\documentclass[%
reprint,
 amsmath,amssymb,
 aps,
]{revtex4-2}
\usepackage{graphicx}
\usepackage{dcolumn}
\usepackage{bm}

\usepackage{color}
\usepackage{orcidlink}
\usepackage{hyperref}
\hypersetup
{
hypertex=true,
colorlinks=true,
linkcolor=blue,
filecolor=blue,
urlcolor=blue,
citecolor=blue,
breaklinks=true
}

\begin{document}

\preprint{APS/123-QED}

\title{A broadband single microwave-photon detector insensitive to the thermal noise}
\author{Y. Q. Chai\orcidlink{0009-0007-6246-9102}}
\affiliation{HergD collaboration and Information Quantum Technology Laboratory, School of Information Science and Technology, Southwest Jiaotong University, Chengdu 610031, China}
\author{M. Y. Wang\orcidlink{0009-0006-6105-950X}}
\affiliation{HergD collaboration and Information Quantum Technology Laboratory, School of Information Science and Technology, Southwest Jiaotong University, Chengdu 610031, China} 
\author{S. N. Wang\orcidlink{0000-0002-5498-4047}}
\affiliation{HergD collaboration and Information Quantum Technology Laboratory, School of Information Science and Technology, Southwest Jiaotong University, Chengdu 610031, China} 
\author{P. H. Ouyang\orcidlink{0009-0000-5379-8552}}
\affiliation{HergD collaboration and Information Quantum Technology Laboratory, School of Information Science and Technology, Southwest Jiaotong University, Chengdu 610031, China}
\author{L. F. Wei\orcidlink{0000-0003-1533-1550}}
\email{ lfwei@swjtu.edu.cn }
\affiliation{HergD collaboration and Information Quantum Technology Laboratory, School of Information Science and Technology, Southwest Jiaotong University, Chengdu 610031, China}

\date{\today}

\begin{abstract}
Thermal noise is one of the physical obstacles that constrain the achievable detection sensitivities of various detectors. Indeed, as we showed in a recent paper (PRB 111, 024501 (2025)), the usual Josephson threshold detector (JTD) operated in an equilibrium state can be utilized to implement a weak microwave signal, just approaching (but not arriving at) its energy quantum limit, even though its physical parameters have been optimized.
In this letter, we further demonstrate numerically that the phase dynamics of a current-biased Josephson junction (CBJJ) can be insensitive to the always-on thermal noise if the sweep rate of the biased current is significantly high. As a consequence, the JTD can be operated alternatively in a non-equilibrium state. Based on the statistical binary detection criterion, we demonstrate how such a non-equilibrium JTD (NEJTD) can be utilized to implement the weak microwave signal, arriving at its energy quantum limit level. The dynamic range and photon-number resolvability of the proposed NEJTD are also discussed when it serves as a broadband single microwave-photon detector.
\end{abstract}
\maketitle

{\it Introduction.---}
Sensitive weak microwave signal detection have played the important roles for various applications, from the wireless communication, radar, sensings to the searches for new physics, etc.~\cite{communication,non-destructive,imaging,electronic1,gravitational1,gravitational2,axion1,axion2,Staggs_2018,computing}. However, how to practically achieve the detection of the weak microwave signal whose energy approaches, even arrives at, its quantum limit (i.e., at the single-photon level) is still a challenge~\cite{ph1,albertinale_detecting_2021}. This is because the achievable sensitivities of both the conventional microwave receivers and the developing single microwave-photon detectors are restricted by the unavoidable thermal noise. 
For example, the achievable detection sensitivity of any traditional room-temperature (at 300~K) microwave receiver is still far from its theoretical upper limit of $-174~{\rm dBm/\sqrt{Hz}}$~\cite{our1}. Various methods with typically the semiconductors~\cite{semiconductor}, graphene devices~\cite{graphene,detector1}, Rydberg atoms~\cite{detector3}, and the superconducting qubits~\cite{JJph3,narrowband1}, etc., have been demonstrated to implement the detections of the itinerant microwave photons; however, the thermal noise induced dark count cannot be avoided practically. This is because the always-on thermal noise cannot be neglected safely, even at temperatures of tens of mK. For example, at 50mK, the upper limit of the achievable detection sensitivity limited by thermal noise is about $-211~{\rm dBm/\sqrt{Hz}}$). This limits the detectability of a single microwave-photon signal, whose energy quantum limit is $\sim 10^{-21}$~mJ. Therefore, the approach to implement the weak signal detection, insensitive to the thermal noise, is particularly desirable.  

Indeed, Josephson junctions, whose plasma frequency is usually in the microwave regime, have been serving as excellent candidates for detecting the microwave photons~\cite{Alesini_2020,detector4,detector5,detector7,JJph3,JJph4,JJph5,PhysRevApplied.8.024022,detector5,graphene_2020}. This is because that, the nonlinear phase dynamics of a CBJJ can be conveniently utilized to implement the threshold detection of weak microwave signals~\cite{detector4,detector7,JJph4,JJph5}, i.e., under the microwave signal driving the CBJJ can be switched from a zero-voltage state to a finite-voltage one~\cite{JJph4,detector7,retrapping,Pankratov2022,detector4,JJph7}. This is a CBJJ-generated Josephson threshold detector (JTD) used for the broadband weak microwave signal detection~\cite{PhysRevB.9.4760}.

However, due to the existence of various random noises, typically including the thermal- and quantum ones, using the JTD to implement the broadband single microwave-photon detection is still a big challenge~\cite{noise1,PhysRevLett.93.106801}. The basic reason is that almost all of the JTDs demonstrated experimentally, up to date, were operated at their equilibrium states, wherein the sweep rates of their linearly-increasing biased currents were sufficiently low. This corresponds to the adiabatic bias of the JTD, and thus the phase dynamics for the used CBJJ is still sensitive to the thermal noise~\cite{our1,our2,detector7,JJph7}. As we showed in Refs.~\cite{our1,our2}, the achievable detection sensitivity of such a JTD can merely approach (e.g., a dozen microwave photons), rather than reach, the desired single microwave photon detection sensitivity, even if the physical parameters of the used CBJJ could be optimized effectively. 

Unlike almost all the previous studies on JTD~\cite{detector4,detector5,detector7,JJph3,JJph4,JJph5,PhysRevApplied.8.024022,detector5,JJph7}, which primarily focus on how to improve detectable sensitivity by optimizing the physical parameters of the device, here we concentrate on how to appropriately set the working parameters of the JTD (typically including the bias-current sweep rate and the initial phase) to relatively mitigate the unavoidable impact of thermal noise, thereby significantly enhancing its achievable detection sensitivity for implementing the single microwave-photon detection. Specifically, by numerically solving the phase dynamics of a noise-driven CBJJ with different bias-current sweep rates and initial phases, we demonstrate that the JTD can operate in a non-equilibrium regime, where its phase dynamics are sensitive to the initial phase but insensitive to thermal noise~\cite{SCD0,nonequilibrium1,nonequilibrium2}. The suppression of thermal noise by non-equilibrium JTD (NEJTD) means an increase in achievable detection sensitivity. Using the statistical distinguishability criterion of binary detection, we numerically demonstrate how to discriminate the switching current distributions (SCD) of the NEJTD with and without a weak microwave signal input. As a consequence, we argue that the NEJTD with the optimized operational parameters can be utilized to implement the highly sensitive detection of weak microwave signals, reaching its energy quantum limit. Interestingly, this is a broadband detection, and naturally, the photon-number resolvability.  

{\it Model and Its Numerical Tests.---} It is well-known that the phase dynamics for the noise-driven CBJJ in a JTD can be described by the following equation~\cite{SCD0,RCSJ3}
\begin{equation}
\begin{aligned}
\frac{{\rm d}^2\varphi}{{\rm d}\tau^2}+\beta\frac{{\rm d}\varphi}{{\rm d}\tau}+\sin(\varphi)=i_b(\tau)+i_n(\tau)\,.
\end{aligned}\label{eq:RCSJ1}
\end{equation}
Here, $\varphi$ is the phase difference between the macroscopic wave functions of the two superconductors across the CBJJ. The noise current $I_n(t)$ and time-dependent bias current $I_b(t)$ have been normalized by the critical current $I_c$ of the junction, yielding the relevant dimensionless quantities read $i_n(\tau)=I_n(t)/I_c$ and $i_b(\tau)=I_b(t)/I_c=v\tau$ with $v$ being the dimensionless sweep rate of the biased current, respectively. Also, $\tau=\omega_Jt$ with $\omega_J=\sqrt{2eI_c/\hbar C}$ being the Plasma frequency of the junction. The dimensionless dissipation coefficient of the CBJJ is represented as $\beta=1/(RC\omega_J)$, with $R$ and $C$ being the resistance and capacitance, respectively. For the simplicity, the noise current is satisfying the relations~\cite{SCD0,thermal}: $\langle i_n(\tau)\rangle=0, \,\langle i_n(\tau)i_n(\tau')\rangle=2\beta k_BT/E_{J0}$, where $E_{J0}=\hbar I_c/2e$ denotes the Josephson energy of the CBJJ, $T$ is the working temperature of the JTD and $k_B$ the Boltzmann constant.

Equation~\eqref{eq:RCSJ1} with $i_n(\tau)=0$ is equivalent to the classical equation of motion for a phase particle with an effective ``mass" $C(\Phi_0/2e)^2$ and ``coordinate" $\varphi$ in the tilted washboard potential $U(\varphi)=E_{J0}\left[1-\cos(\varphi)-i_b(\tau)\varphi\right],\,i_b(\tau)<1$. For $i_b(\tau)<1$, the phase particle remains trapped in a potential well of height $\Delta U(\varphi)=2E_{J0}\left[\sqrt{1-i_b(\tau)^2}-i_b(\tau)\arccos(i_b(\tau))\right]$, and the CBJJ stays in the zero-voltage state. As $i_b(\tau)$ increases toward unity, the barrier is progressively suppressed until the particle escapes, giving rise to a measurable finite voltage across the CBJJ. However, due to the existence of noise current $i_n(\tau)$, a finite probability of the CBJJ can be switched from the zero-voltage state to a non-zero one, even for certain biased current $i_b(\tau)=i_{\rm{sw}}<1$~\cite{noise1,noise2}. Consequently, an SCD could be obtained by the experimentally repeated measurements of the switching currents $\{i_{\rm{sw}}\}$. If the biased current is slowly increased with the low sweep rate $v$, the measured SCD can be fitted by using the usual semi-empirical formula~\cite{SCD_eq}  
\begin{equation}
\begin{aligned}
P(i_{sw})=\frac{\Gamma(i_{sw})}{v}\int_0^{i_{sw}}v\Gamma(i'_{sw})di'_{sw}\,,
\end{aligned} \label{eq:SCD}
\end{equation}
where $\Gamma(i_{sw})$ is the phase particle escape probability caused by thermal activation~\cite{noise2} and quantum tunneling~\cite{Leggett}. Therefore, the detection sensitivity of an equilibrium JTD (EJTD) with low sweep rate $v$ is inevitably limited by thermal noise at typical operating temperatures $T \geq T_{\rm cr}\simeq54$~mK~\cite{JJph4}, where thermal noise dominates. Our previous work has shown that merely optimizing the physical parameters of the CBJJ allows the EJTD to approach, but not reach, the energy quantum limit for weak microwave signal detection.

In practice, the operating temperature of a conventional EJTD cannot be further reduced, making the mitigation of thermal noise the only viable route to enhance its sensitivity for weak microwave signal detection. To verify this, we numerically solve Eq.~\eqref{eq:RCSJ1} for different bias-current sweep rates and initial phases.
\begin{figure}[htbp]
\includegraphics[width=1\linewidth]{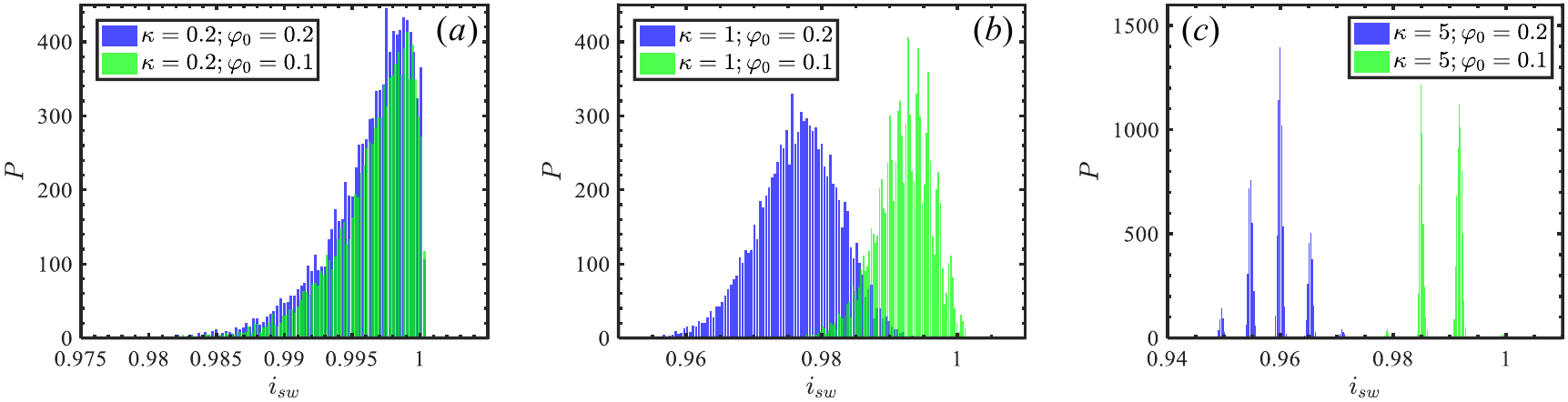}
\caption{Numerical simulations of the SCDs for a JTD biased by the currents with the typical sweep rates. Here, the relevant parameters are set as:  $\dot{\varphi}_{0}=0$, $\beta=10^{-4}$, $v=1\times10^{-5}$, and $2\beta k_{B}T/E_{J0}=1\times10^{-7}$ which corresponds to the typical thermal noise level of a microampere-scale JJ around 50~mK, with a plasma frequency of several GHz.}\label{F1}
\end{figure}
For a given working temperature with $2\beta k_{B}T/E_{J0}=1\times10^{-7}$ and the fixed $\beta$-parameter, the corresponding SCDs are shown in Fig.~\ref{F1}, wherein an effective sweep-rate parameter $\kappa=v/\beta$ is introduced to characterize the bias-current sweep rates. It is seen that the phase dynamics of the JTD depend sensitively on the value of the $\kappa$-parameter, which delivers two distinct working regimes of the JTD. In the adiabatically biased regime, i.e., $\kappa \ll 1$, the JTD is operated in its equilibrium state, wherein the phase dynamics are insensitive to the initial phase. This behavior is confirmed by the SCD obtained from $10^4$ numerical solutions of Eq.~\eqref{eq:RCSJ1} and shows the excellent agreement with experimental data~\cite{RCSJ3}. 
In contrast, for the non-adiabatically biased currents with $\kappa \geq 1$, the JTD is operated in its non-equilibrium state~\cite{nonequilibrium1,nonequilibrium2}. The corresponding SCDs, shown in Fig.~\ref{F1}(b, c), exhibit the manifest dependence of the initial phase, even in the presence of thermal noise.
These results show that the SCD variation induced by the initial phase serves as a fingerprint of the NEJTD’s insensitivity to thermal noise.

Therefore, by identifying whether the initial phase fingerprint of the SCD for NEJTD operating at different temperatures changes, one can check whether the phase dynamics of NEJTD are sensitive to thermal noise or not. Fortunately, the receiver operating characteristic curve (AUC) in statistics and its area parameter $\mathcal{R}_{\rm AUC}$~\cite{ROC2,ROC3,classify} can be directly used to measure the distinguishability of two arbitrary SCDs; the larger value of the $\mathcal{R}_{\rm AUC}$-parameter corresponds to the higher distinguishability. Certainly, if the value of this parameter is 0.5, it indicates that the two statistical distributions are completely indistinguishable.

\begin{figure}[htbp]
\includegraphics[width=0.9\linewidth]{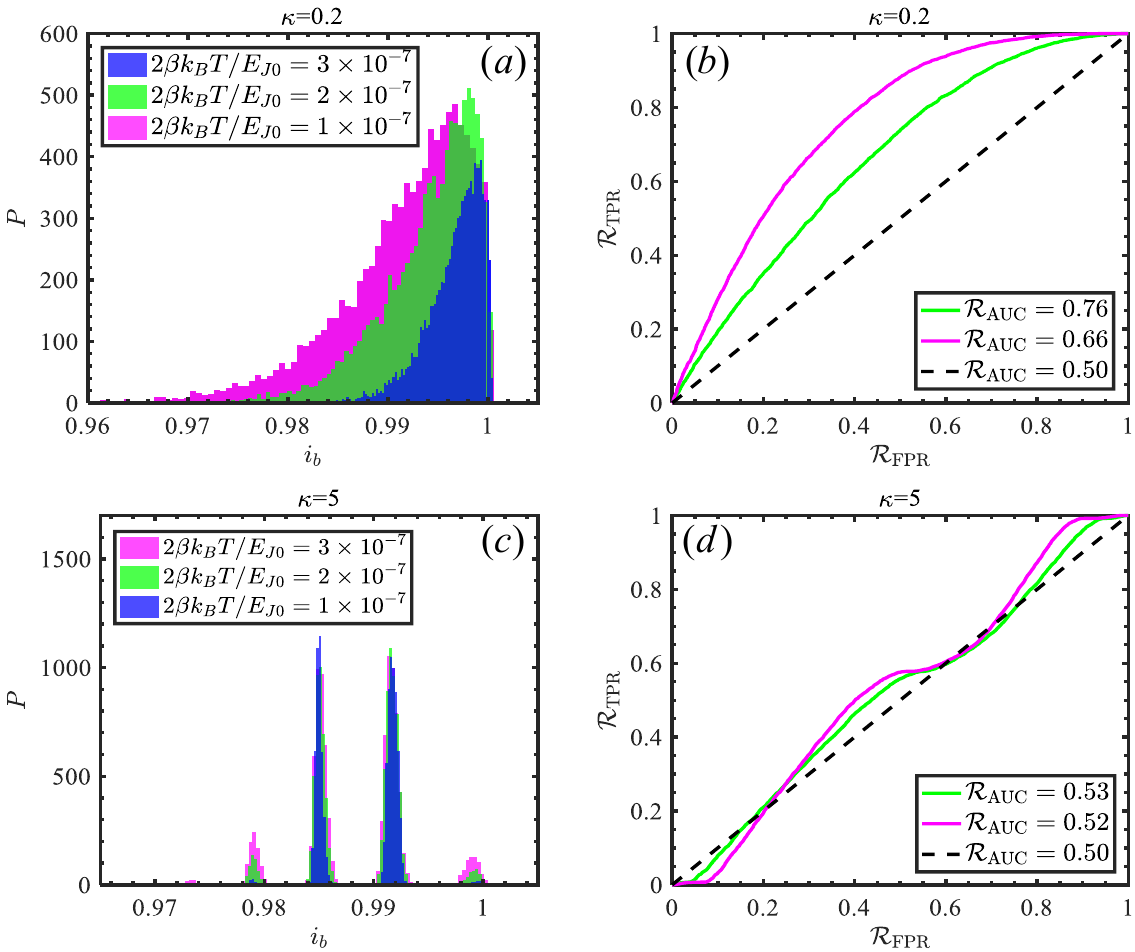}
\caption{Numerical verification of the thermal-noise sensitivity in the SCDs with $\varphi_{0}=0.1$; (a, b) for the EJTD with $\kappa=0.2$, and (c, d) for the NEJTD with $\kappa=5$. The parameters are the same as in Fig.~\ref{F1}.}\label{F2}
\end{figure}
Figs.~\ref{F2}(a, c) show the SCDs of the EJTD and NEJTD, for the different thermal noise levels. It is seen from Fig.~\ref{F2}(a) that the SCDs of the usual EJTD remain unimodal across the temperatures, yielding a distinguishability of $\mathcal{R}_{\rm AUC}=0.76>0.66>0.5$ for the ROC curve shown in Fig.~\ref{F2}(b). This is the manifest dependence of the thermal noise. In contrast, for the NEJTD, one can see from Fig.~\ref{F2}(c) that the SCDs lose their unimodal characters, and the ROC curve shown in Fig.~\ref{F2}(d) gives $\mathcal{R}_{\rm AUC}=0.53\approx0.52\approx0.5$, demonstrating that the influence of the variation of thermal noise on its SCD has been greatly suppressed. This confirms that the phase dynamics of the NEJTD are manifestly insensitive to the thermal noise, although it is always-on.

Physically, the dissipation factor $\beta$ arises from thermal noise acting on the phase particles, while the rapidly swept bias current serves as an external driving force. As the sweep rate $\kappa$ increases, the influence of thermal noise on the NEJTD phase dynamics diminishes. Compared to the EJTD, the NEJTD’s reduced sensitivity to thermal noise makes it promising for single microwave-photon detection.

\begin{figure}[htbp]
\includegraphics[width=0.9\linewidth]{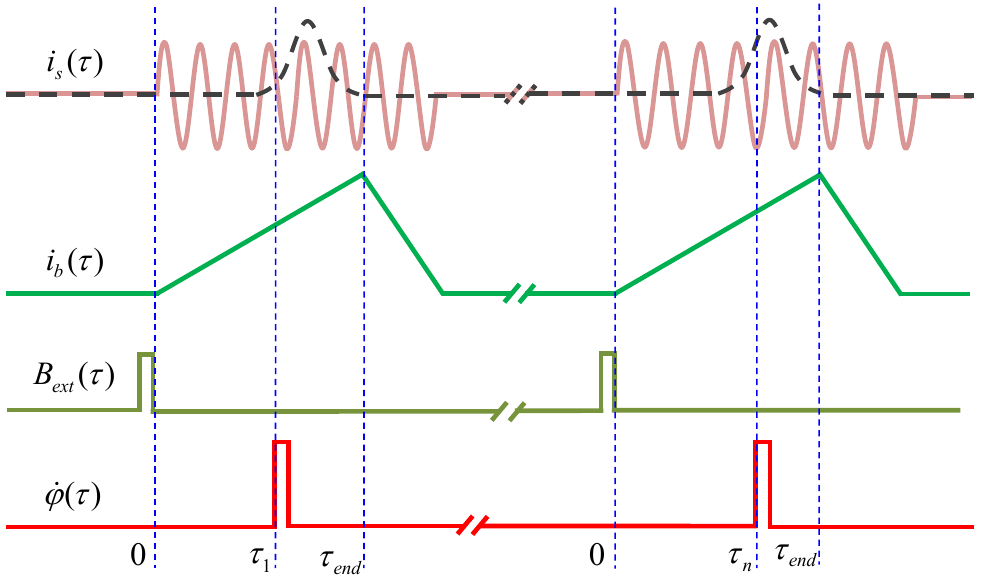}
\caption{Schematic of the weak microwave signal detection scheme by using a NEJTD. From top to bottom: incident microwave signal, triangular bias current, external magnetic field for phase initialization, and real-time voltage response of the JTD.}\label{F3}
\end{figure}

{\it Method and its Numerical Verifications.---}
The numerical results demonstrated above suggest that the NEJTD can be generated to implement the highly sensitive detection of weak microwave signals. In fact, the physical model, by using a JTD to implement the detection of a weak microwave signal $i_s(\tau)$, can be generically described by the following equation~\cite{detector4,JJph4,JJph7,our1,our2}
\begin{equation}
\frac{{\rm d}^2\varphi}{{\rm d}\tau^2}+\beta\frac{{\rm d}\varphi}{{\rm d}\tau}+\sin(\varphi)=v\tau+i_n(\tau)+i_s(\tau)\,.
\label{eq:detection}
\end{equation}
Here, we set $\kappa=v/\beta\geq 1$ for driving the JTD in its non-equilibrium state, where its phase dynamics are insensitive to noise but sensitive to its initial phase. 
Specifically, the scheme, by using a JTD to implement the detection of the continuous (pale red)/pulsed (black) microwave signal $i_s(\tau)$, can be schematically shown in Fig.~\ref{F3}. Here, a rapidly swept biased current (green), which increases linearly until a voltage switch occurs, is applied to drive the JTD working in its non-equilibrium regime. A magnetic field $B_{\rm ext}(\tau)$ (olive) is used to synchronically modulate the initial phase of the JTD, whose amplitude is kept well below the critical magnetic field to avoid any unwanted breaking of Cooper-pair tunnelings. Within each bias cycle, the time zero is defined as the onset of the triangular drive, and the $n$-th switching current is measured when the $n$-th appearance of the voltage signal. Consequently, the SCDs with and without the input of signal $i_s(\tau)$ can be obtained. 
Obviously, the present method to implement the weak microwave signal detection departs from the conventional schemes using the usual EJTDs for the same aim in two essential ways. First, the bias current is swept at a sufficiently high rate, so that the JTD is operated at its non-equilibrium state for effectively suppressing the influence of the thermal noise. Secondly, a tunable DC bias is applied to generate an adjustable magnetic field, enabling the initial phase of the JTD to be flexibly set. This makes the measured SCDs characterized by the initial-phase fingerprints, which are insensitive to thermal noise. Detection of the input signal is then achieved by monitoring the changes in these SCD fingerprints.
With the operational scheme shown in Fig.~\ref{F3}, the detection of the microwave signal $i_s(\tau)$ can be performed sequentially as the follows: i) first, the SCD of the JTD without the signal inputs, i.e., $i_s(\tau)=0$, are measured and it is marked as $P_0=(SCD)_0$; ii) the SCD of the JTD with the signal inputs, i.e., $i_s(\tau)\neq 0$, are measured and it is marked as $P_1=(SCD)_1$; and iii) the judgment of whether there is a microwave signal input can be given, by realizing the distinction between $P_1$ and $P_0$, on the data processing side.
Below, we demonstrate its feasibility by numerical experiments.

\begin{figure}[htbp]
\includegraphics[width=1\linewidth]{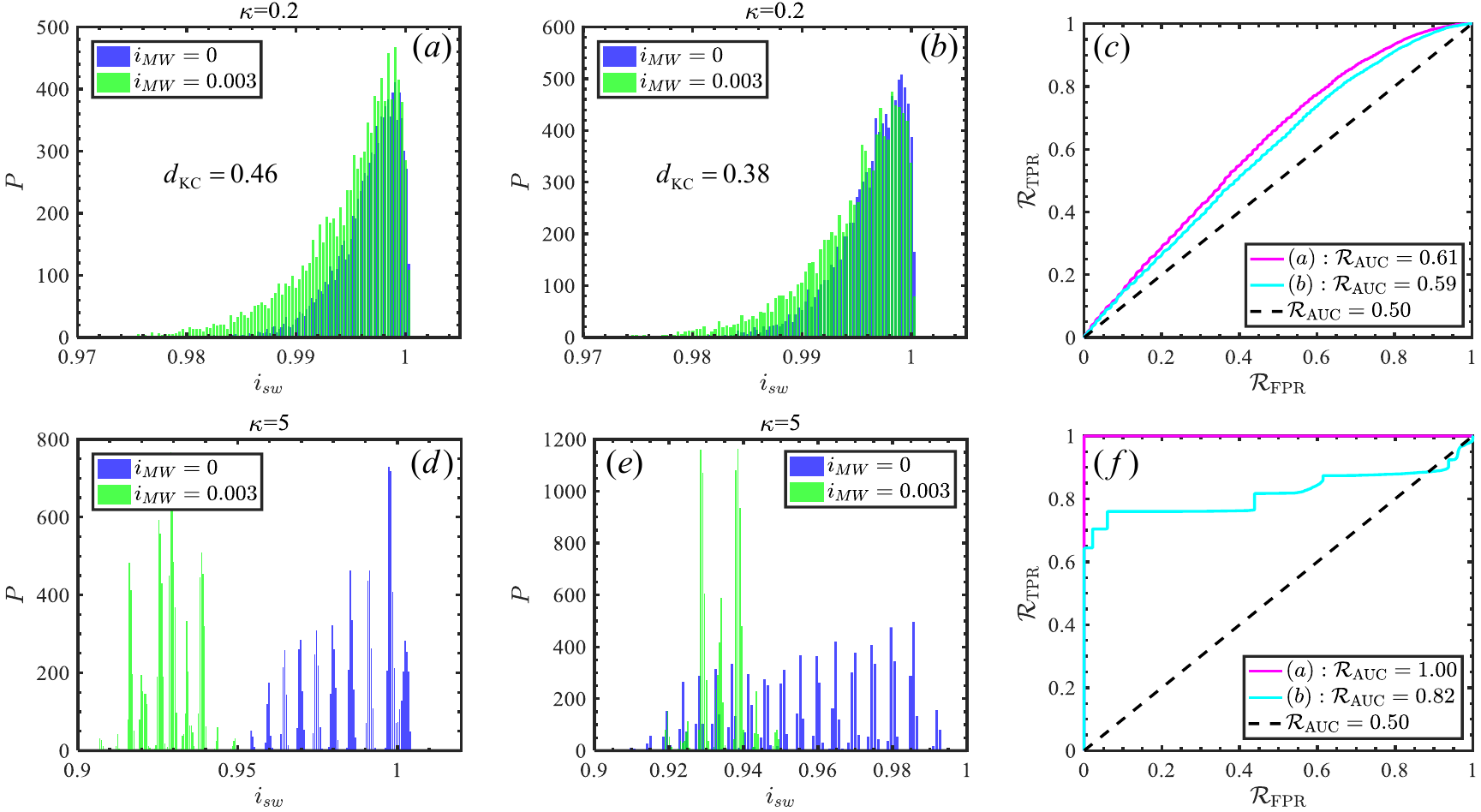}
\caption{Simulated distinguishability of the SCDs with and without the continuous-wave signal input, for different initial phases. (a, b) for an EJTD, while (d,e) for a NEJTD. (c, f) show the corresponding distinguishabilities. Here, the relevant parameters are set as: $\omega_{MW}=1$, $\varphi_0=0.1$ in (a, d) and $\varphi_0=0.2$ in (b, e). The other parameters as in Fig.~\ref{F1}.}\label{F4}
\end{figure}

First, let us consider the detection of a weak continuous-wave microwave signal represented as 
\begin{equation}
i_s(\tau)=i_{MW}\sin(\omega_{MW}\tau),
\end{equation}
with $i_{MW}$ and $\omega_{MW}$ being the amplitude and frequency, normalized to $I_c$ and $\omega_J$, respectively. 
For the comparison, Figs.~\ref{F4}(a, b) show the SCDs of the usual EJTD biased by the current $i_b(\tau)$ with the typically low $\kappa$-parameter, with and without the detected signal input specifically for $i_{MW}=0.003$. It is seen that the values of $d_{KC}$-index are slightly greater than the indistinguishable $\min[d_{KC}]$~\cite{our1}, which means that the detectability of the signal by using such an EJTD is very low. The corresponding ROC curve and its $\mathcal{R}_ {\rm AUC} $-parameter values shown in Fig.~\ref{F4} (c) are $0.59$ and $0.61$ respectively, presenting similar results. Their distinguishability is very poor, showing that the influence of the thermal noise cannot be neglected. 
Very differently, if the JTD is operated in its non-equilibrium state, typically with $\kappa=5$, the SCDs shown in Figs.~\ref{F4}(d, e), with and without the signal input, possess the obvious initial fingerprints. Their ROC curves shown in Fig.~\ref{F4}(f) indicates that the corresponding values of the distinguishable $\mathcal{R}_{\rm AUC}$-parameter can be enhanced as $\mathcal{R}_{\rm AUC}=0.82$ for $\varphi_0=0.1$ and $\mathcal{R}_{\rm AUC}=1.00$ for $\varphi_0=0.2$, respectively. This clearly demonstrates that, in the same noise environment, the NEJTD possesses a stronger weak signal detection ability, compared with its equilibrium counterpart, as the noise impacts in the NEJTD had been isolated effectively. 

\begin{figure}[htbp]
\includegraphics[width=1\linewidth]{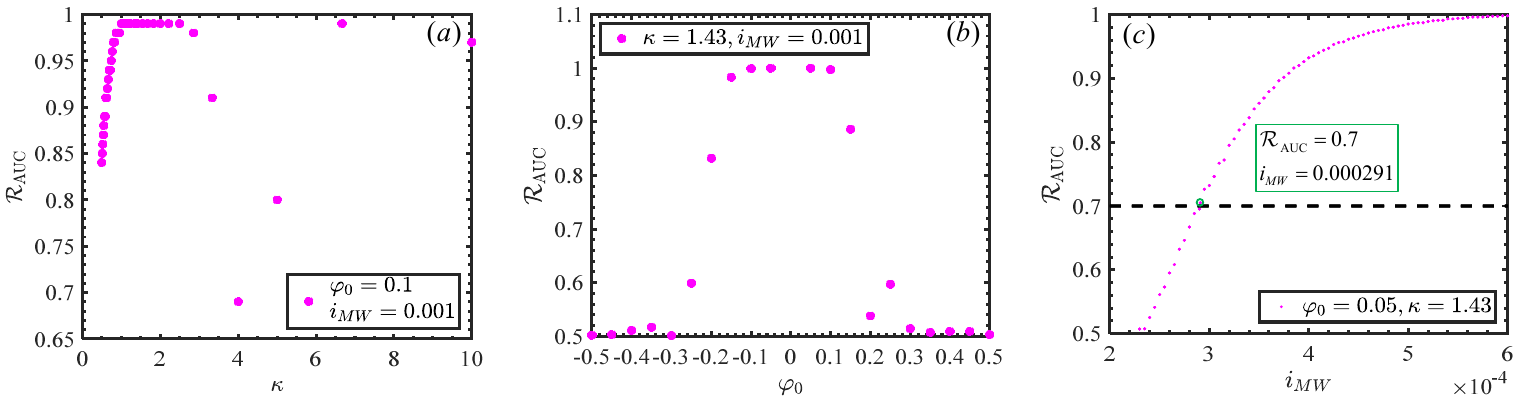}
\caption{Numerical optimization of the operational parameters of a NEJTD for single-photon signal detection. (a) $\kappa$ parameter, (b) $\varphi_0$ parameter, and (c) signal intensity. Here, $\omega_{MW}=1$ and the other parameters as in Fig.~\ref{F1}.}\label{F5}
\end{figure}

Now, we estimate the achievable detection sensitivity of an NEJTD by optimizing its adjustably operational parameters. Specifically, Fig.~\ref{F5} illustrates their optimized results; Fig.~\ref{F5}(a) shows the optimized sweep rate ($\kappa$-parameter), for the fixed $\varphi_0=0.1$. While, Fig.~\ref{F5}(b) shows the optimized initial phase $\varphi_0$ for the optimized sweep rate $\kappa=1.43$. With these optimized operational parameters, the detectability of the optimized NEJTD for the input weak microwave signals is shown in Fig.~\ref{F5}(c), and demonstrating that the value of the $\mathcal{R}_{\rm AUC}$-parameter increases monotonically with the signal intensity.
According to the statistical binary detection criterion~\cite{ROC3}, i.e., the reliable detection is ensured once $\mathcal{R}_{\rm AUC}\geq\mathcal{R}^T_{\rm AUC}=0.7$, the weakest detectable signal of the optimized NEJTD can be estimated as $i_{MW}=2.91\times10^{-4}$. Accordingly, the minimum detectable power can be expressed as~\cite{PhysRevApplied.22.024015}: $P_{min}=i_{MW}^2I_c^2R_{MW}/(2\chi)\approx8.4681\times10^{-6}I_c^2$\,, for $R_{MW}=100~\Omega$ and $\chi=0.5$, under the ideal impedance matching. As the critical current $I_c$ of the CBJJ in JTD can be reduced to the few-nanoampere range~\cite{detector7}, the minimum detectable microwave signal power, by using the optimized NEJTD, could reach the order of $10^{-23}~{\rm J/s}$, which is close to the energy quantum limit of the microwave signal at a unit bandwidth. 

\begin{figure}[htbp]
\includegraphics[width=1\linewidth]{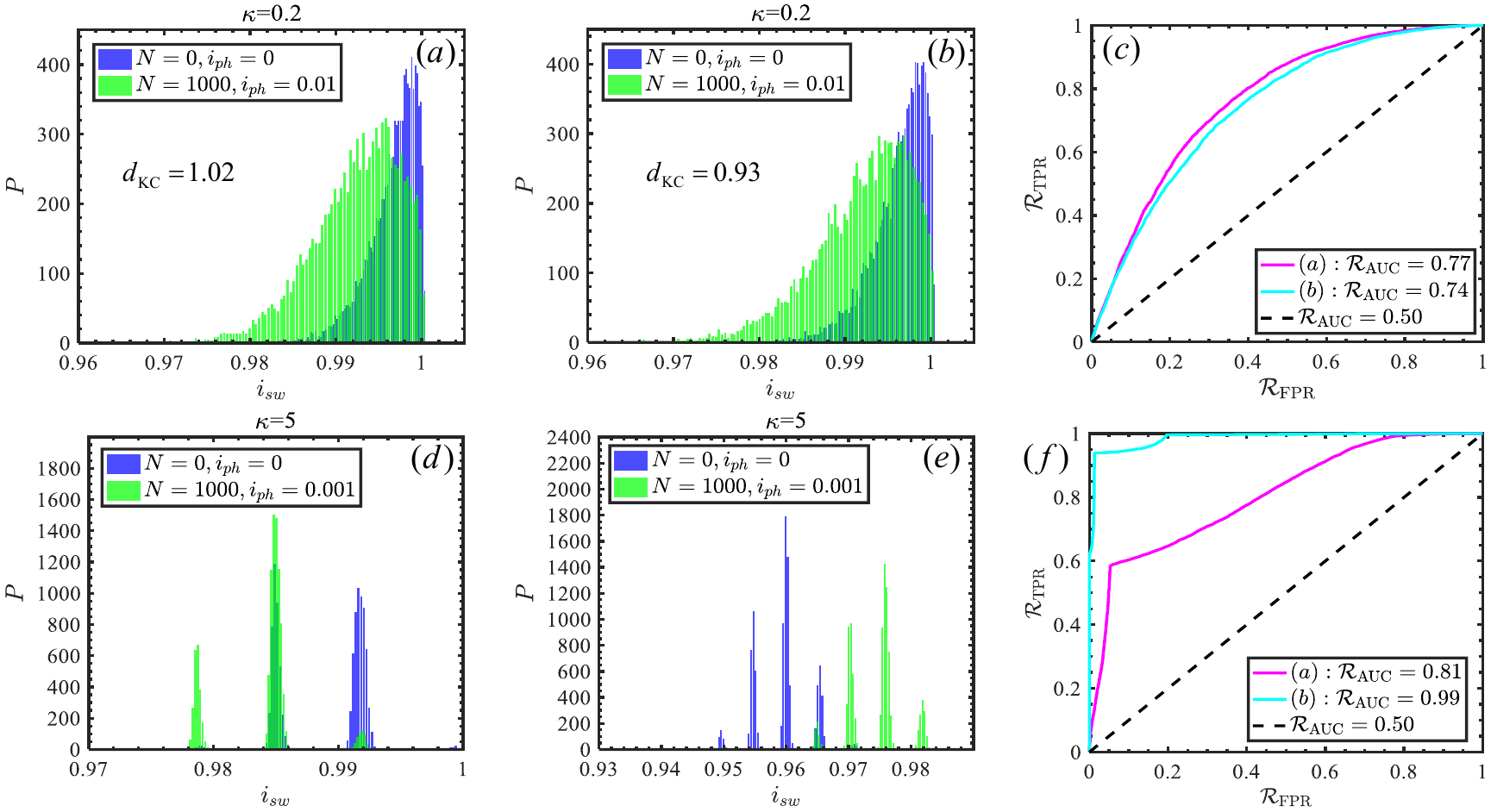}
\caption{Numerical simulations of the SCDs for a weak pulsed microwave signal; (a, b) is for an EJTD, and (d, e) is for an NEJTD. (c, f) show the corresponding detectabilities. The pulse parameters are set as: $\omega_{ph}=1$, $\tau_{ph}=1$, $\tau_d=1/(2v)=1/(2\kappa\beta)=25000$. The other parameters as in Fig.~\ref{F1} and Fig.~\ref{F4}.}\label{F6}
\end{figure}

Next, we specifically demonstrate the detectability of a microwave-photon pulsed signal (i.e., the black line shown in Fig.~\ref{F3}) by using the above NEJTD. Following Ref.~\cite{pulse}, the applied pulsed microwave current can be expressed formally as
\begin{equation}
\begin{aligned}
i_{ph}(\omega_{ph}, \tau_{ ph}, \tau)=&\sqrt{N_{ph}}i_{ph}\exp\left(-\frac{1}{2}(\frac{\tau-\tau_{d}}{\tau_{ph}})^2\right)\\
&\times\cos(\omega_{ph}(\tau-\tau_{d}))\,,\label{eq:ph}
\end{aligned}
\end{equation}
in terms of the photon number. Here, $N_{ph}$ is the number of photons contained in the pulse. Also, $\tau_{ph}$ and $\tau_{d}$ are the width and arrival time of the pulse, respectively. The signal frequency $\omega_{ph}$ has been normalized by the Josephson plasma frequency $\omega_J$. Also, the amplitude of the pulsed microwave current has been normalized by the critical current $I_c$ of the junction with the normalized amplitude, i.e., $i_{ph}=I_{ ph}/I_c=\sqrt{\hbar\omega_{ph}\omega_J^{2}/(RI_c^{2}\tau_{ph})}$,  
is determined by the critical current $I_c$ of the junction. Specifically, for a pulse signal containing $N=1000$ photons, Fig.~\ref{F6}(a, b) shows the SCDs obtained by using the EJTDs with $\kappa=0.2$ for $\varphi_0=0.1$ and $0.2$, respectively. It is seen that, although the corresponding values of $d_{KC}$-index indicate their distinguishability, the ROC curves shown in Fig.~\ref{F6}(c) yield the relatively low $\mathcal{R}_{\rm AUC}$ values, which are only slightly above the threshold $\mathcal{R}^T_{\rm AUC}=0.7$. This again evidences the very limited ability of the EJTD.
By contrast, for the NEJTD typically with $\kappa=5$, Figs.~\ref{F6}(d, e) show that their SCDs display the pronounced fingerprints of initial phases, and the corresponding ROC curves shown in Fig.~\ref{F6}(f) yield $\mathcal{R}_{\rm AUC}=0.81$ and $0.99$ for $\varphi_0=0.1$ and $0.2$, respectively. These results establish the manifestly enhanced detection capability of the NEJTDs for the weak microwave pulsed signal under the same conditions.

\begin{figure}[htbp]
\includegraphics[width=1\linewidth]{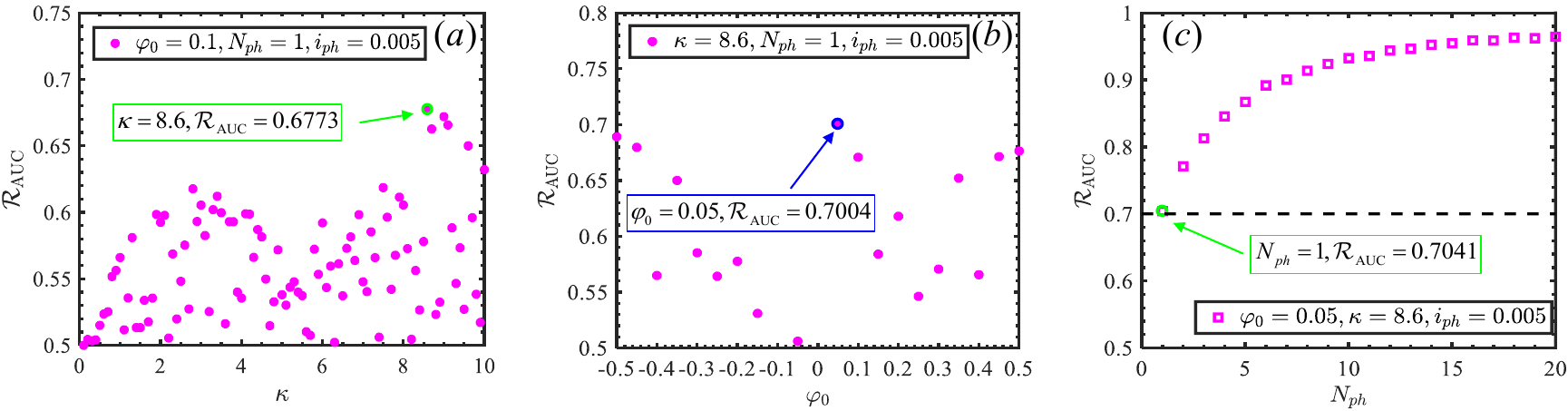}
\caption{Numerical optimizations  of the operational parameters for an NEJTD; (a) for $\kappa$-parameter and (b) for $\varphi_0$-parameter. (c) Its detectability for the pulsed signal with different photon numbers. Here, the pulsed parameters are set as: $\omega_{ph}=1$, $\tau_{ph}=356$ (5 ns), and $\tau_d=1/(2v)=1/(2\kappa\beta)$. The other parameters as in Fig.~\ref{F1}.}\label{F7}
\end{figure}

Given that the achievable detection of the EJTDs with the optimized physical parameters can approach the energy quantum limit of a pulsed microwave signal (i.e., a few photons level)~\cite{our1}, it is therefore reasonable to expect that, by further optimizing its operational parameters, the NEJTD can be implementing the more sensitive detection of a pulsed microwave signal, arriving at single microwave-photon level. With the optimized operational parameters shown specifically in Figs.~\ref{F7}(a, b), we numerically demonstrate the detectability of the NEJTD, with optimized operational parameters (i.e., $\kappa=8.6$ and $\varphi_0=0.05$), for the pulsed microwave signals containing different numbers of photons. Remarkably, by using the such a NEJTD, a single microwave-photon pulsed signal, with $N_{ph}=1$ and $\tau\sim 5$~ns and thus  $i_{ph}^{\min}=I_{ph}/I_c\approx5.7~\text{nA}/0.975~\mu\text{A}\approx0.005$~\cite{our1,our3}, can still be detection relizably, as its $\mathcal{R}_{\rm AUC}$-parameter is still larger that the reliable detection threshold of $\mathcal{R}_{\rm AUC}^T=0.7$~\cite{ROC3}. 
Furthermore, the numerical results shown in Fig.~\ref{F7}(c) also provide the dynamic range of the NEJTD, i.e., the response remains approximately linear~\cite{9769720} for the increasing signal strength. Within this regime, the value of the $\mathcal{R}_{\rm AUC}$-parameter corresponds to the photon number contained in the pulsed microwave signal, one-to-one. The onset of nonlinearity determines the upper bound of the detectability. The detectable curve shown as the dotted line in Fig.~\ref{F7}(c) clearly represents that, such as a detector possesses naturally the ability for implementing the photon-number resolution and the maximal resolvable photon number being $N_{ph}^{\max}=15$. The other achievable performance indexes, such as the detection quantum efficiency, dead duration, and dark count rate, can be similarly investigated by the relevant numerical simulations.

\begin{figure}[htbp]
\includegraphics[width=1\linewidth]{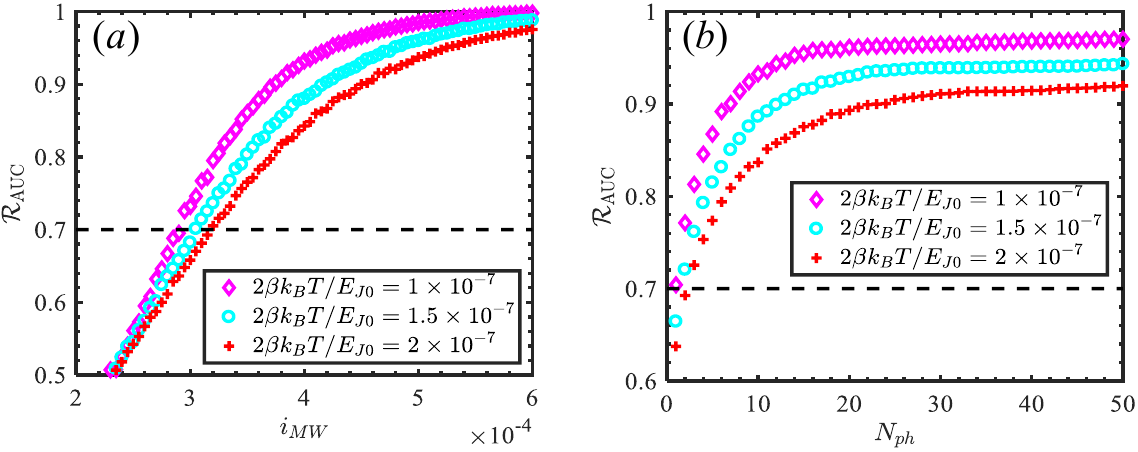}
\caption{The detectability of the optimized NEJTD influenced by the thermal noises; (a) for the continuous-wave signal, and (b) for the pulsed one .}\label{F8}
\end{figure}

To further confirm the capability of the NEJTD detector for the weak microwave-photon signal, which is really insensitive to the thermal noise, in Fig.~\ref{F8} we show how its $\mathcal{R}_{\rm AUC}$-parameter changes are influenced by the thermal noise, for both the weak continuous-wave signal and pulsed one, respectively. The results show that the impacts of thermal noises are really very weak, as demonstrated previously in Fig.~\ref{F2}. Even when the noise intensity is doubled, the detector still possesses a sufficiently high achievable detection sensitivity for the photon-number resolvable detection of a weak microwave-photon signal, at its energy quantum limit. 

{\it Conclusion.---} In summary, we demonstrated, by numerical methods, that the phase dynamics of the CBJJ biased by current with sufficiently high sweep rate, can be insensitive to the thermal noise. Therefore, it can be used to generate NEJTD for the desired single microwave-photon detection. Since the impact of thermal noise can be effectively isolated, this detector has extremely high sensitivity and naturally possesses the photon-number-resolution ability. 

As the physical parameters of the device and its operational parameters, used for the numerical simulations, are achievable at the current experimental level, we believe that the proposed detector is feasible, although the fundamental detection mechanism still merits further investigations. Moreover, how to use the proposed detector to achieve the desired single-shot detection of a single microwave-photon pulsed signal, rather than the statistical detection presented here, might be the next theoretical and experimental challenge that needs to be addressed. These unresolved problems will be discussed elsewhere. 

{\it Acknowledgments.---} This work is partially supported by the National Key Research and Development Program of China (NKRDC) under Grant No. 2021YFA0718803, and the National Natural Science Foundation of China (NSFC) under Grant No. 11974290.
\bibliography{Reference}

\begin{thebibliography}{49}%
\makeatletter
\providecommand \@ifxundefined [1]{%
 \@ifx{#1\undefined}
}%
\providecommand \@ifnum [1]{%
 \ifnum #1\expandafter \@firstoftwo
 \else \expandafter \@secondoftwo
 \fi
}%
\providecommand \@ifx [1]{%
 \ifx #1\expandafter \@firstoftwo
 \else \expandafter \@secondoftwo
 \fi
}%
\providecommand \natexlab [1]{#1}%
\providecommand \enquote  [1]{``#1''}%
\providecommand \bibnamefont  [1]{#1}%
\providecommand \bibfnamefont [1]{#1}%
\providecommand \citenamefont [1]{#1}%
\providecommand \href@noop [0]{\@secondoftwo}%
\providecommand \href [0]{\begingroup \@sanitize@url \@href}%
\providecommand \@href[1]{\@@startlink{#1}\@@href}%
\providecommand \@@href[1]{\endgroup#1\@@endlink}%
\providecommand \@sanitize@url [0]{\catcode `\\12\catcode `\$12\catcode
  `\&12\catcode `\#12\catcode `\^12\catcode `\_12\catcode `\%12\relax}%
\providecommand \@@startlink[1]{}%
\providecommand \@@endlink[0]{}%
\providecommand \url  [0]{\begingroup\@sanitize@url \@url }%
\providecommand \@url [1]{\endgroup\@href {#1}{\urlprefix }}%
\providecommand \urlprefix  [0]{URL }%
\providecommand \Eprint [0]{\href }%
\providecommand \doibase [0]{https://doi.org/}%
\providecommand \selectlanguage [0]{\@gobble}%
\providecommand \bibinfo  [0]{\@secondoftwo}%
\providecommand \bibfield  [0]{\@secondoftwo}%
\providecommand \translation [1]{[#1]}%
\providecommand \BibitemOpen [0]{}%
\providecommand \bibitemStop [0]{}%
\providecommand \bibitemNoStop [0]{.\EOS\space}%
\providecommand \EOS [0]{\spacefactor3000\relax}%
\providecommand \BibitemShut  [1]{\csname bibitem#1\endcsname}%
\let\auto@bib@innerbib\@empty
\bibitem [{\citenamefont {Deng}\ \emph {et~al.}(2024)\citenamefont {Deng},
  \citenamefont {Burasa},\ and\ \citenamefont {Wu}}]{communication}%
  \BibitemOpen
  \bibfield  {author} {\bibinfo {author} {\bibfnamefont {J.}~\bibnamefont
  {Deng}}, \bibinfo {author} {\bibfnamefont {P.}~\bibnamefont {Burasa}},\ and\
  \bibinfo {author} {\bibfnamefont {K.}~\bibnamefont {Wu}},\ }\bibfield
  {title} {\bibinfo {title} {Joint multiband linear interferometric receiver
  for integrated microwave and terahertz sensing and communication systems},\
  }\href {https://doi.org/10.1109/TMTT.2024.3363173} {\bibfield  {journal}
  {\bibinfo  {journal} {IEEE Trans. Microw. Theory Tech.}\ }\textbf {\bibinfo
  {volume} {72}},\ \bibinfo {pages} {5550} (\bibinfo {year}
  {2024})}\BibitemShut {NoStop}%
\bibitem [{\citenamefont {Brinker}\ \emph {et~al.}(2020)\citenamefont
  {Brinker}, \citenamefont {Dvorsky}, \citenamefont {Al~Qaseer},\ and\
  \citenamefont {Zoughi}}]{non-destructive}%
  \BibitemOpen
  \bibfield  {author} {\bibinfo {author} {\bibfnamefont {K.}~\bibnamefont
  {Brinker}}, \bibinfo {author} {\bibfnamefont {M.}~\bibnamefont {Dvorsky}},
  \bibinfo {author} {\bibfnamefont {M.~T.}\ \bibnamefont {Al~Qaseer}},\ and\
  \bibinfo {author} {\bibfnamefont {R.}~\bibnamefont {Zoughi}},\ }\bibfield
  {title} {\bibinfo {title} {Review of advances in microwave and
  millimetre-wave ndt\&e: Principles and applications},\ }\href
  {https://doi.org/https://doi.org/10.1098/rsta.2019.0585} {\bibfield
  {journal} {\bibinfo  {journal} {Phil. Trans. R. Soc. A.}\ }\textbf {\bibinfo
  {volume} {378}},\ \bibinfo {pages} {20190585} (\bibinfo {year}
  {2020})}\BibitemShut {NoStop}%
\bibitem [{\citenamefont {Chandra}\ \emph {et~al.}(2015)\citenamefont
  {Chandra}, \citenamefont {Zhou}, \citenamefont {Balasingham},\ and\
  \citenamefont {Narayanan}}]{imaging}%
  \BibitemOpen
  \bibfield  {author} {\bibinfo {author} {\bibfnamefont {R.}~\bibnamefont
  {Chandra}}, \bibinfo {author} {\bibfnamefont {H.}~\bibnamefont {Zhou}},
  \bibinfo {author} {\bibfnamefont {I.}~\bibnamefont {Balasingham}},\ and\
  \bibinfo {author} {\bibfnamefont {R.~M.}\ \bibnamefont {Narayanan}},\
  }\bibfield  {title} {\bibinfo {title} {On the opportunities and challenges in
  microwave medical sensing and imaging},\ }\href
  {https://doi.org/10.1109/TBME.2015.2432137} {\bibfield  {journal} {\bibinfo
  {journal} {IEEE Trans. Biomed. Eng.}\ }\textbf {\bibinfo {volume} {62}},\
  \bibinfo {pages} {1667} (\bibinfo {year} {2015})}\BibitemShut {NoStop}%
\bibitem [{\citenamefont {Jia}\ \emph {et~al.}(2024)\citenamefont {Jia},
  \citenamefont {Wang}, \citenamefont {Song}, \citenamefont {Cui},
  \citenamefont {Chen}, \citenamefont {Wang}, \citenamefont {Liu},
  \citenamefont {Zhang},\ and\ \citenamefont {Bao}}]{electronic1}%
  \BibitemOpen
  \bibfield  {author} {\bibinfo {author} {\bibfnamefont {L.}~\bibnamefont
  {Jia}}, \bibinfo {author} {\bibfnamefont {Y.}~\bibnamefont {Wang}}, \bibinfo
  {author} {\bibfnamefont {Y.}~\bibnamefont {Song}}, \bibinfo {author}
  {\bibfnamefont {W.}~\bibnamefont {Cui}}, \bibinfo {author} {\bibfnamefont
  {Z.}~\bibnamefont {Chen}}, \bibinfo {author} {\bibfnamefont {R.}~\bibnamefont
  {Wang}}, \bibinfo {author} {\bibfnamefont {Y.}~\bibnamefont {Liu}}, \bibinfo
  {author} {\bibfnamefont {W.}~\bibnamefont {Zhang}},\ and\ \bibinfo {author}
  {\bibfnamefont {M.}~\bibnamefont {Bao}},\ }\bibfield  {title} {\bibinfo
  {title} {The detection technology of high-power microwave: A review},\ }\href
  {https://doi.org/10.1109/TIM.2024.3472802} {\bibfield  {journal} {\bibinfo
  {journal} {IEEE Trans. Instrum. Meas.}\ }\textbf {\bibinfo {volume} {73}},\
  \bibinfo {pages} {1} (\bibinfo {year} {2024})}\BibitemShut {NoStop}%
\bibitem [{\citenamefont {Berlin}\ \emph {et~al.}(2022)\citenamefont {Berlin},
  \citenamefont {Blas}, \citenamefont {D'Agnolo}, \citenamefont {Ellis},
  \citenamefont {Harnik}, \citenamefont {Kahn},\ and\ \citenamefont
  {Sch\"utte-Engel}}]{gravitational1}%
  \BibitemOpen
  \bibfield  {author} {\bibinfo {author} {\bibfnamefont {A.}~\bibnamefont
  {Berlin}}, \bibinfo {author} {\bibfnamefont {D.}~\bibnamefont {Blas}},
  \bibinfo {author} {\bibfnamefont {R.~T.}\ \bibnamefont {D'Agnolo}}, \bibinfo
  {author} {\bibfnamefont {S.~A.~R.}\ \bibnamefont {Ellis}}, \bibinfo {author}
  {\bibfnamefont {R.}~\bibnamefont {Harnik}}, \bibinfo {author} {\bibfnamefont
  {Y.}~\bibnamefont {Kahn}},\ and\ \bibinfo {author} {\bibfnamefont
  {J.}~\bibnamefont {Sch\"utte-Engel}},\ }\bibfield  {title} {\bibinfo {title}
  {Detecting high-frequency gravitational waves with microwave cavities},\
  }\href {https://doi.org/10.1103/PhysRevD.105.116011} {\bibfield  {journal}
  {\bibinfo  {journal} {Phys. Rev. D}\ }\textbf {\bibinfo {volume} {105}},\
  \bibinfo {pages} {116011} (\bibinfo {year} {2022})}\BibitemShut {NoStop}%
\bibitem [{\citenamefont {Zheng}\ \emph {et~al.}(2018)\citenamefont {Zheng},
  \citenamefont {Wei}, \citenamefont {Wen},\ and\ \citenamefont
  {Li}}]{gravitational2}%
  \BibitemOpen
  \bibfield  {author} {\bibinfo {author} {\bibfnamefont {H.}~\bibnamefont
  {Zheng}}, \bibinfo {author} {\bibfnamefont {L.~F.}\ \bibnamefont {Wei}},
  \bibinfo {author} {\bibfnamefont {H.}~\bibnamefont {Wen}},\ and\ \bibinfo
  {author} {\bibfnamefont {F.~Y.}\ \bibnamefont {Li}},\ }\bibfield  {title}
  {\bibinfo {title} {Electromagnetic response of gravitational waves passing
  through an alternating magnetic field: A scheme to probe high-frequency
  gravitational waves},\ }\href {https://doi.org/10.1103/PhysRevD.98.064028}
  {\bibfield  {journal} {\bibinfo  {journal} {Phys. Rev. D}\ }\textbf {\bibinfo
  {volume} {98}},\ \bibinfo {pages} {064028} (\bibinfo {year}
  {2018})}\BibitemShut {NoStop}%
\bibitem [{\citenamefont {Bradley}\ \emph {et~al.}(2003)\citenamefont
  {Bradley}, \citenamefont {Clarke}, \citenamefont {Kinion}, \citenamefont
  {Rosenberg}, \citenamefont {van Bibber}, \citenamefont {Matsuki},
  \citenamefont {M\"uck},\ and\ \citenamefont {Sikivie}}]{axion1}%
  \BibitemOpen
  \bibfield  {author} {\bibinfo {author} {\bibfnamefont {R.}~\bibnamefont
  {Bradley}}, \bibinfo {author} {\bibfnamefont {J.}~\bibnamefont {Clarke}},
  \bibinfo {author} {\bibfnamefont {D.}~\bibnamefont {Kinion}}, \bibinfo
  {author} {\bibfnamefont {L.~J.}\ \bibnamefont {Rosenberg}}, \bibinfo {author}
  {\bibfnamefont {K.}~\bibnamefont {van Bibber}}, \bibinfo {author}
  {\bibfnamefont {S.}~\bibnamefont {Matsuki}}, \bibinfo {author} {\bibfnamefont
  {M.}~\bibnamefont {M\"uck}},\ and\ \bibinfo {author} {\bibfnamefont
  {P.}~\bibnamefont {Sikivie}},\ }\bibfield  {title} {\bibinfo {title}
  {Microwave cavity searches for dark-matter axions},\ }\href
  {https://doi.org/10.1103/RevModPhys.75.777} {\bibfield  {journal} {\bibinfo
  {journal} {Rev. Mod. Phys.}\ }\textbf {\bibinfo {volume} {75}},\ \bibinfo
  {pages} {777} (\bibinfo {year} {2003})}\BibitemShut {NoStop}%
\bibitem [{\citenamefont {Lasenby}(2021)}]{axion2}%
  \BibitemOpen
  \bibfield  {author} {\bibinfo {author} {\bibfnamefont {R.}~\bibnamefont
  {Lasenby}},\ }\bibfield  {title} {\bibinfo {title} {Parametrics of
  electromagnetic searches for axion dark matter},\ }\href
  {https://doi.org/10.1103/PhysRevD.103.075007} {\bibfield  {journal} {\bibinfo
   {journal} {Phys. Rev. D}\ }\textbf {\bibinfo {volume} {103}},\ \bibinfo
  {pages} {075007} (\bibinfo {year} {2021})}\BibitemShut {NoStop}%
\bibitem [{\citenamefont {Staggs}\ \emph {et~al.}(2018)\citenamefont {Staggs},
  \citenamefont {Dunkley},\ and\ \citenamefont {Page}}]{Staggs_2018}%
  \BibitemOpen
  \bibfield  {author} {\bibinfo {author} {\bibfnamefont {S.}~\bibnamefont
  {Staggs}}, \bibinfo {author} {\bibfnamefont {J.}~\bibnamefont {Dunkley}},\
  and\ \bibinfo {author} {\bibfnamefont {L.}~\bibnamefont {Page}},\ }\bibfield
  {title} {\bibinfo {title} {Recent discoveries from the cosmic microwave
  background: a review of recent progress},\ }\href
  {https://doi.org/10.1088/1361-6633/aa94d5} {\bibfield  {journal} {\bibinfo
  {journal} {Rep. Prog. Phys.}\ }\textbf {\bibinfo {volume} {81}},\ \bibinfo
  {pages} {044901} (\bibinfo {year} {2018})}\BibitemShut {NoStop}%
\bibitem [{\citenamefont {Senanian}\ \emph {et~al.}(2024)\citenamefont
  {Senanian}, \citenamefont {Prabhu}, \citenamefont {Kremenetski},
  \citenamefont {Roy}, \citenamefont {Cao}, \citenamefont {Kline},
  \citenamefont {Onodera}, \citenamefont {Wright}, \citenamefont {Wu},
  \citenamefont {Fatemi} \emph {et~al.}}]{computing}%
  \BibitemOpen
  \bibfield  {author} {\bibinfo {author} {\bibfnamefont {A.}~\bibnamefont
  {Senanian}}, \bibinfo {author} {\bibfnamefont {S.}~\bibnamefont {Prabhu}},
  \bibinfo {author} {\bibfnamefont {V.}~\bibnamefont {Kremenetski}}, \bibinfo
  {author} {\bibfnamefont {S.}~\bibnamefont {Roy}}, \bibinfo {author}
  {\bibfnamefont {Y.}~\bibnamefont {Cao}}, \bibinfo {author} {\bibfnamefont
  {J.}~\bibnamefont {Kline}}, \bibinfo {author} {\bibfnamefont
  {T.}~\bibnamefont {Onodera}}, \bibinfo {author} {\bibfnamefont {L.~G.}\
  \bibnamefont {Wright}}, \bibinfo {author} {\bibfnamefont {X.}~\bibnamefont
  {Wu}}, \bibinfo {author} {\bibfnamefont {V.}~\bibnamefont {Fatemi}}, \emph
  {et~al.},\ }\bibfield  {title} {\bibinfo {title} {Microwave signal processing
  using an analog quantum reservoir computer},\ }\href
  {https://doi.org/10.1038/s41467-024-51161-8} {\bibfield  {journal} {\bibinfo
  {journal} {Nat. Commun.}\ }\textbf {\bibinfo {volume} {15}},\ \bibinfo
  {pages} {7490} (\bibinfo {year} {2024})}\BibitemShut {NoStop}%
\bibitem [{\citenamefont {Il’ichev}(2016)}]{ph1}%
  \BibitemOpen
  \bibfield  {author} {\bibinfo {author} {\bibfnamefont {E.}~\bibnamefont
  {Il’ichev}},\ }\bibfield  {title} {\bibinfo {title} {A microwave photon
  detector},\ }\href {https://doi.org/10.1134/S1063783416110123} {\bibfield
  {journal} {\bibinfo  {journal} {Phys. Solid State}\ }\textbf {\bibinfo
  {volume} {58}},\ \bibinfo {pages} {2160} (\bibinfo {year}
  {2016})}\BibitemShut {NoStop}%
\bibitem [{\citenamefont {Albertinale}\ \emph {et~al.}()\citenamefont
  {Albertinale}, \citenamefont {Balembois}, \citenamefont {Billaud},
  \citenamefont {Ranjan}, \citenamefont {Flanigan}, \citenamefont {Schenkel},
  \citenamefont {Estève}, \citenamefont {Vion}, \citenamefont {Bertet},\ and\
  \citenamefont {Flurin}}]{albertinale_detecting_2021}%
  \BibitemOpen
  \bibfield  {author} {\bibinfo {author} {\bibfnamefont {E.}~\bibnamefont
  {Albertinale}}, \bibinfo {author} {\bibfnamefont {L.}~\bibnamefont
  {Balembois}}, \bibinfo {author} {\bibfnamefont {E.}~\bibnamefont {Billaud}},
  \bibinfo {author} {\bibfnamefont {V.}~\bibnamefont {Ranjan}}, \bibinfo
  {author} {\bibfnamefont {D.}~\bibnamefont {Flanigan}}, \bibinfo {author}
  {\bibfnamefont {T.}~\bibnamefont {Schenkel}}, \bibinfo {author}
  {\bibfnamefont {D.}~\bibnamefont {Estève}}, \bibinfo {author} {\bibfnamefont
  {D.}~\bibnamefont {Vion}}, \bibinfo {author} {\bibfnamefont {P.}~\bibnamefont
  {Bertet}},\ and\ \bibinfo {author} {\bibfnamefont {E.}~\bibnamefont
  {Flurin}},\ }\bibfield  {title} {\bibinfo {title} {Detecting spins by their
  fluorescence with a microwave photon counter},\ }\href
  {https://doi.org/10.1038/s41586-021-04076-z} {\bibfield  {journal} {\bibinfo
  {journal} {Nature}\ }\textbf {\bibinfo {volume} {600}},\ \bibinfo {pages}
  {434}}\BibitemShut {NoStop}%
\bibitem [{\citenamefont {Chai}\ \emph {et~al.}(2025)\citenamefont {Chai},
  \citenamefont {Wang}, \citenamefont {OuYang},\ and\ \citenamefont
  {Wei}}]{our1}%
  \BibitemOpen
  \bibfield  {author} {\bibinfo {author} {\bibfnamefont {Y.~Q.}\ \bibnamefont
  {Chai}}, \bibinfo {author} {\bibfnamefont {S.~N.}\ \bibnamefont {Wang}},
  \bibinfo {author} {\bibfnamefont {P.~H.}\ \bibnamefont {OuYang}},\ and\
  \bibinfo {author} {\bibfnamefont {L.~F.}\ \bibnamefont {Wei}},\ }\bibfield
  {title} {\bibinfo {title} {Measuring weak microwave signals via
  current-biased josephson junctions: Approaching the quantum limit of energy
  detection},\ }\href {https://doi.org/10.1103/PhysRevB.111.024501} {\bibfield
  {journal} {\bibinfo  {journal} {Phys. Rev. B}\ }\textbf {\bibinfo {volume}
  {111}},\ \bibinfo {pages} {024501} (\bibinfo {year} {2025})}\BibitemShut
  {NoStop}%
\bibitem [{\citenamefont {Cornia}\ \emph {et~al.}(2023)\citenamefont {Cornia},
  \citenamefont {Demontis}, \citenamefont {Zannier}, \citenamefont {Sorba},
  \citenamefont {Ghirri}, \citenamefont {Rossella},\ and\ \citenamefont
  {Affronte}}]{semiconductor}%
  \BibitemOpen
  \bibfield  {author} {\bibinfo {author} {\bibfnamefont {S.}~\bibnamefont
  {Cornia}}, \bibinfo {author} {\bibfnamefont {V.}~\bibnamefont {Demontis}},
  \bibinfo {author} {\bibfnamefont {V.}~\bibnamefont {Zannier}}, \bibinfo
  {author} {\bibfnamefont {L.}~\bibnamefont {Sorba}}, \bibinfo {author}
  {\bibfnamefont {A.}~\bibnamefont {Ghirri}}, \bibinfo {author} {\bibfnamefont
  {F.}~\bibnamefont {Rossella}},\ and\ \bibinfo {author} {\bibfnamefont
  {M.}~\bibnamefont {Affronte}},\ }\bibfield  {title} {\bibinfo {title}
  {Calibration-free and high-sensitivity microwave detectors based on inas/inp
  nanowire double quantum dots},\ }\href
  {https://doi.org/https://doi.org/10.1002/adfm.202212517} {\bibfield
  {journal} {\bibinfo  {journal} {Adv. Funct. Mater.}\ }\textbf {\bibinfo
  {volume} {33}},\ \bibinfo {pages} {2212517} (\bibinfo {year}
  {2023})}\BibitemShut {NoStop}%
\bibitem [{\citenamefont {Saeed}\ \emph {et~al.}(2022)\citenamefont {Saeed},
  \citenamefont {Palacios}, \citenamefont {Wei}, \citenamefont {Baskent},
  \citenamefont {Fan}, \citenamefont {Uzlu}, \citenamefont {Wang},
  \citenamefont {Hemmetter}, \citenamefont {Wang}, \citenamefont {Neumaier},
  \citenamefont {Lemme},\ and\ \citenamefont {Negra}}]{graphene}%
  \BibitemOpen
  \bibfield  {author} {\bibinfo {author} {\bibfnamefont {M.}~\bibnamefont
  {Saeed}}, \bibinfo {author} {\bibfnamefont {P.}~\bibnamefont {Palacios}},
  \bibinfo {author} {\bibfnamefont {M.-D.}\ \bibnamefont {Wei}}, \bibinfo
  {author} {\bibfnamefont {E.}~\bibnamefont {Baskent}}, \bibinfo {author}
  {\bibfnamefont {C.-Y.}\ \bibnamefont {Fan}}, \bibinfo {author} {\bibfnamefont
  {B.}~\bibnamefont {Uzlu}}, \bibinfo {author} {\bibfnamefont {K.-T.}\
  \bibnamefont {Wang}}, \bibinfo {author} {\bibfnamefont {A.}~\bibnamefont
  {Hemmetter}}, \bibinfo {author} {\bibfnamefont {Z.}~\bibnamefont {Wang}},
  \bibinfo {author} {\bibfnamefont {D.}~\bibnamefont {Neumaier}}, \bibinfo
  {author} {\bibfnamefont {M.~C.}\ \bibnamefont {Lemme}},\ and\ \bibinfo
  {author} {\bibfnamefont {R.}~\bibnamefont {Negra}},\ }\bibfield  {title}
  {\bibinfo {title} {Graphene-based microwave circuits: A review},\ }\href
  {https://doi.org/https://doi.org/10.1002/adma.202108473} {\bibfield
  {journal} {\bibinfo  {journal} {Adv. Mater.}\ }\textbf {\bibinfo {volume}
  {34}},\ \bibinfo {pages} {2108473} (\bibinfo {year} {2022})}\BibitemShut
  {NoStop}%
\bibitem [{\citenamefont {Westlund}\ \emph {et~al.}(2015)\citenamefont
  {Westlund}, \citenamefont {Winters}, \citenamefont {Ivanov}, \citenamefont
  {Hassan}, \citenamefont {Nilsson}, \citenamefont {Janzén}, \citenamefont
  {Rorsman},\ and\ \citenamefont {Grahn}}]{detector1}%
  \BibitemOpen
  \bibfield  {author} {\bibinfo {author} {\bibfnamefont {A.}~\bibnamefont
  {Westlund}}, \bibinfo {author} {\bibfnamefont {M.}~\bibnamefont {Winters}},
  \bibinfo {author} {\bibfnamefont {I.~G.}\ \bibnamefont {Ivanov}}, \bibinfo
  {author} {\bibfnamefont {J.}~\bibnamefont {Hassan}}, \bibinfo {author}
  {\bibfnamefont {P.-{\AA}.}\ \bibnamefont {Nilsson}}, \bibinfo {author}
  {\bibfnamefont {E.}~\bibnamefont {Janzén}}, \bibinfo {author} {\bibfnamefont
  {N.}~\bibnamefont {Rorsman}},\ and\ \bibinfo {author} {\bibfnamefont
  {J.}~\bibnamefont {Grahn}},\ }\bibfield  {title} {\bibinfo {title} {Graphene
  self-switching diodes as zero-bias microwave detectors},\ }\href
  {https://doi.org/10.1063/1.4914356} {\bibfield  {journal} {\bibinfo
  {journal} {Appl. Phys. Lett.}\ }\textbf {\bibinfo {volume} {106}},\ \bibinfo
  {pages} {093116} (\bibinfo {year} {2015})}\BibitemShut {NoStop}%
\bibitem [{\citenamefont {Jing}\ \emph {et~al.}(2020)\citenamefont {Jing},
  \citenamefont {Hu}, \citenamefont {Ma}, \citenamefont {Zhang}, \citenamefont
  {Zhang}, \citenamefont {Xiao},\ and\ \citenamefont {Jia}}]{detector3}%
  \BibitemOpen
  \bibfield  {author} {\bibinfo {author} {\bibfnamefont {M.}~\bibnamefont
  {Jing}}, \bibinfo {author} {\bibfnamefont {Y.}~\bibnamefont {Hu}}, \bibinfo
  {author} {\bibfnamefont {J.}~\bibnamefont {Ma}}, \bibinfo {author}
  {\bibfnamefont {H.}~\bibnamefont {Zhang}}, \bibinfo {author} {\bibfnamefont
  {L.}~\bibnamefont {Zhang}}, \bibinfo {author} {\bibfnamefont
  {L.}~\bibnamefont {Xiao}},\ and\ \bibinfo {author} {\bibfnamefont
  {S.}~\bibnamefont {Jia}},\ }\bibfield  {title} {\bibinfo {title} {Atomic
  superheterodyne receiver based on microwave-dressed rydberg spectroscopy},\
  }\href {https://doi.org/10.1038/s41567-020-0918-5} {\bibfield  {journal}
  {\bibinfo  {journal} {Nat. Phys.}\ }\textbf {\bibinfo {volume} {16}},\
  \bibinfo {pages} {911} (\bibinfo {year} {2020})}\BibitemShut {NoStop}%
\bibitem [{\citenamefont {Poudel}\ \emph {et~al.}(2012)\citenamefont {Poudel},
  \citenamefont {McDermott},\ and\ \citenamefont {Vavilov}}]{JJph3}%
  \BibitemOpen
  \bibfield  {author} {\bibinfo {author} {\bibfnamefont {A.}~\bibnamefont
  {Poudel}}, \bibinfo {author} {\bibfnamefont {R.}~\bibnamefont {McDermott}},\
  and\ \bibinfo {author} {\bibfnamefont {M.~G.}\ \bibnamefont {Vavilov}},\
  }\bibfield  {title} {\bibinfo {title} {Quantum efficiency of a microwave
  photon detector based on a current-biased josephson junction},\ }\href
  {https://doi.org/10.1103/PhysRevB.86.174506} {\bibfield  {journal} {\bibinfo
  {journal} {Phys. Rev. B}\ }\textbf {\bibinfo {volume} {86}},\ \bibinfo
  {pages} {174506} (\bibinfo {year} {2012})}\BibitemShut {NoStop}%
\bibitem [{\citenamefont {Chen}\ \emph {et~al.}(2011)\citenamefont {Chen},
  \citenamefont {Hover}, \citenamefont {Sendelbach}, \citenamefont {Maurer},
  \citenamefont {Merkel}, \citenamefont {Pritchett}, \citenamefont {Wilhelm},\
  and\ \citenamefont {McDermott}}]{narrowband1}%
  \BibitemOpen
  \bibfield  {author} {\bibinfo {author} {\bibfnamefont {Y.-F.}\ \bibnamefont
  {Chen}}, \bibinfo {author} {\bibfnamefont {D.}~\bibnamefont {Hover}},
  \bibinfo {author} {\bibfnamefont {S.}~\bibnamefont {Sendelbach}}, \bibinfo
  {author} {\bibfnamefont {L.}~\bibnamefont {Maurer}}, \bibinfo {author}
  {\bibfnamefont {S.~T.}\ \bibnamefont {Merkel}}, \bibinfo {author}
  {\bibfnamefont {E.~J.}\ \bibnamefont {Pritchett}}, \bibinfo {author}
  {\bibfnamefont {F.~K.}\ \bibnamefont {Wilhelm}},\ and\ \bibinfo {author}
  {\bibfnamefont {R.}~\bibnamefont {McDermott}},\ }\bibfield  {title} {\bibinfo
  {title} {Microwave photon counter based on josephson junctions},\ }\href
  {https://doi.org/10.1103/PhysRevLett.107.217401} {\bibfield  {journal}
  {\bibinfo  {journal} {Phys. Rev. Lett.}\ }\textbf {\bibinfo {volume} {107}},\
  \bibinfo {pages} {217401} (\bibinfo {year} {2011})}\BibitemShut {NoStop}%
\bibitem [{\citenamefont {Alesini}\ \emph {et~al.}(2020)\citenamefont
  {Alesini}, \citenamefont {Babusci}, \citenamefont {Barone}, \citenamefont
  {Buonomo}, \citenamefont {Beretta}, \citenamefont {Bianchini}, \citenamefont
  {Castellano}, \citenamefont {Chiarello}, \citenamefont {Di~Gioacchino},
  \citenamefont {Falferi}, \citenamefont {Felici}, \citenamefont {Filatrella},
  \citenamefont {Foggetta}, \citenamefont {Gallo}, \citenamefont {Gatti},
  \citenamefont {Giazotto}, \citenamefont {Lamanna}, \citenamefont {Ligabue},
  \citenamefont {Ligato}, \citenamefont {Ligi}, \citenamefont {Maccarrone},
  \citenamefont {Margesin}, \citenamefont {Mattioli}, \citenamefont
  {Monticone}, \citenamefont {Oberto}, \citenamefont {Pagano}, \citenamefont
  {Paolucci}, \citenamefont {Rajteri}, \citenamefont {Rettaroli}, \citenamefont
  {Rolandi}, \citenamefont {Spagnolo}, \citenamefont {Toncelli},\ and\
  \citenamefont {Torrioli}}]{Alesini_2020}%
  \BibitemOpen
  \bibfield  {author} {\bibinfo {author} {\bibfnamefont {D.}~\bibnamefont
  {Alesini}}, \bibinfo {author} {\bibfnamefont {D.}~\bibnamefont {Babusci}},
  \bibinfo {author} {\bibfnamefont {C.}~\bibnamefont {Barone}}, \bibinfo
  {author} {\bibfnamefont {B.}~\bibnamefont {Buonomo}}, \bibinfo {author}
  {\bibfnamefont {M.~M.}\ \bibnamefont {Beretta}}, \bibinfo {author}
  {\bibfnamefont {L.}~\bibnamefont {Bianchini}}, \bibinfo {author}
  {\bibfnamefont {G.}~\bibnamefont {Castellano}}, \bibinfo {author}
  {\bibfnamefont {F.}~\bibnamefont {Chiarello}}, \bibinfo {author}
  {\bibfnamefont {D.}~\bibnamefont {Di~Gioacchino}}, \bibinfo {author}
  {\bibfnamefont {P.}~\bibnamefont {Falferi}}, \bibinfo {author} {\bibfnamefont
  {G.}~\bibnamefont {Felici}}, \bibinfo {author} {\bibfnamefont
  {G.}~\bibnamefont {Filatrella}}, \bibinfo {author} {\bibfnamefont {L.~G.}\
  \bibnamefont {Foggetta}}, \bibinfo {author} {\bibfnamefont {A.}~\bibnamefont
  {Gallo}}, \bibinfo {author} {\bibfnamefont {C.}~\bibnamefont {Gatti}},
  \bibinfo {author} {\bibfnamefont {F.}~\bibnamefont {Giazotto}}, \bibinfo
  {author} {\bibfnamefont {G.}~\bibnamefont {Lamanna}}, \bibinfo {author}
  {\bibfnamefont {F.}~\bibnamefont {Ligabue}}, \bibinfo {author} {\bibfnamefont
  {N.}~\bibnamefont {Ligato}}, \bibinfo {author} {\bibfnamefont
  {C.}~\bibnamefont {Ligi}}, \bibinfo {author} {\bibfnamefont {G.}~\bibnamefont
  {Maccarrone}}, \bibinfo {author} {\bibfnamefont {B.}~\bibnamefont
  {Margesin}}, \bibinfo {author} {\bibfnamefont {F.}~\bibnamefont {Mattioli}},
  \bibinfo {author} {\bibfnamefont {E.}~\bibnamefont {Monticone}}, \bibinfo
  {author} {\bibfnamefont {L.}~\bibnamefont {Oberto}}, \bibinfo {author}
  {\bibfnamefont {S.}~\bibnamefont {Pagano}}, \bibinfo {author} {\bibfnamefont
  {F.}~\bibnamefont {Paolucci}}, \bibinfo {author} {\bibfnamefont
  {M.}~\bibnamefont {Rajteri}}, \bibinfo {author} {\bibfnamefont
  {A.}~\bibnamefont {Rettaroli}}, \bibinfo {author} {\bibfnamefont
  {L.}~\bibnamefont {Rolandi}}, \bibinfo {author} {\bibfnamefont
  {P.}~\bibnamefont {Spagnolo}}, \bibinfo {author} {\bibfnamefont
  {A.}~\bibnamefont {Toncelli}},\ and\ \bibinfo {author} {\bibfnamefont
  {G.}~\bibnamefont {Torrioli}},\ }\bibfield  {title} {\bibinfo {title}
  {Development of a josephson junction based single photon microwave detector
  for axion detection experiments},\ }\href
  {https://doi.org/10.1088/1742-6596/1559/1/012020} {\bibfield  {journal}
  {\bibinfo  {journal} {J. Phys.: Conf. Ser.}\ }\textbf {\bibinfo {volume}
  {1559}},\ \bibinfo {pages} {012020} (\bibinfo {year} {2020})}\BibitemShut
  {NoStop}%
\bibitem [{\citenamefont {Revin}\ \emph {et~al.}(2020)\citenamefont {Revin},
  \citenamefont {Pankratov}, \citenamefont {Gordeeva}, \citenamefont
  {Yablokov}, \citenamefont {Rakut}, \citenamefont {Zbrozhek},\ and\
  \citenamefont {Kuzmin}}]{detector4}%
  \BibitemOpen
  \bibfield  {author} {\bibinfo {author} {\bibfnamefont {L.~S.}\ \bibnamefont
  {Revin}}, \bibinfo {author} {\bibfnamefont {A.~L.}\ \bibnamefont
  {Pankratov}}, \bibinfo {author} {\bibfnamefont {A.~V.}\ \bibnamefont
  {Gordeeva}}, \bibinfo {author} {\bibfnamefont {A.~A.}\ \bibnamefont
  {Yablokov}}, \bibinfo {author} {\bibfnamefont {I.~V.}\ \bibnamefont {Rakut}},
  \bibinfo {author} {\bibfnamefont {V.~O.}\ \bibnamefont {Zbrozhek}},\ and\
  \bibinfo {author} {\bibfnamefont {L.~S.}\ \bibnamefont {Kuzmin}},\ }\bibfield
   {title} {\bibinfo {title} {Microwave photon detection by an al josephson
  junction},\ }\href {https://doi.org/10.3762/bjnano.11.80} {\bibfield
  {journal} {\bibinfo  {journal} {Beilstein J. Nanotechnol.}\ }\textbf
  {\bibinfo {volume} {11}},\ \bibinfo {pages} {960} (\bibinfo {year}
  {2020})}\BibitemShut {NoStop}%
\bibitem [{\citenamefont {Kokkoniemi}\ \emph {et~al.}(2020)\citenamefont
  {Kokkoniemi}, \citenamefont {Girard}, \citenamefont {Hazra}, \citenamefont
  {Laitinen}, \citenamefont {Govenius}, \citenamefont {Lake}, \citenamefont
  {Sallinen}, \citenamefont {Vesterinen}, \citenamefont {Partanen},
  \citenamefont {Tan} \emph {et~al.}}]{detector5}%
  \BibitemOpen
  \bibfield  {author} {\bibinfo {author} {\bibfnamefont {R.}~\bibnamefont
  {Kokkoniemi}}, \bibinfo {author} {\bibfnamefont {J.-P.}\ \bibnamefont
  {Girard}}, \bibinfo {author} {\bibfnamefont {D.}~\bibnamefont {Hazra}},
  \bibinfo {author} {\bibfnamefont {A.}~\bibnamefont {Laitinen}}, \bibinfo
  {author} {\bibfnamefont {J.}~\bibnamefont {Govenius}}, \bibinfo {author}
  {\bibfnamefont {R.}~\bibnamefont {Lake}}, \bibinfo {author} {\bibfnamefont
  {I.}~\bibnamefont {Sallinen}}, \bibinfo {author} {\bibfnamefont
  {V.}~\bibnamefont {Vesterinen}}, \bibinfo {author} {\bibfnamefont
  {M.}~\bibnamefont {Partanen}}, \bibinfo {author} {\bibfnamefont
  {J.}~\bibnamefont {Tan}}, \emph {et~al.},\ }\bibfield  {title} {\bibinfo
  {title} {Bolometer operating at the threshold for circuit quantum
  electrodynamics},\ }\href {https://doi.org/10.1038/s41586-020-2753-3}
  {\bibfield  {journal} {\bibinfo  {journal} {Nature}\ }\textbf {\bibinfo
  {volume} {586}},\ \bibinfo {pages} {47} (\bibinfo {year} {2020})}\BibitemShut
  {NoStop}%
\bibitem [{\citenamefont {Pankratov}\ \emph
  {et~al.}(2022{\natexlab{a}})\citenamefont {Pankratov}, \citenamefont {Revin},
  \citenamefont {Gordeeva}, \citenamefont {Yablokov}, \citenamefont {Kuzmin},\
  and\ \citenamefont {Il’Ichev}}]{detector7}%
  \BibitemOpen
  \bibfield  {author} {\bibinfo {author} {\bibfnamefont {A.}~\bibnamefont
  {Pankratov}}, \bibinfo {author} {\bibfnamefont {L.}~\bibnamefont {Revin}},
  \bibinfo {author} {\bibfnamefont {A.}~\bibnamefont {Gordeeva}}, \bibinfo
  {author} {\bibfnamefont {A.}~\bibnamefont {Yablokov}}, \bibinfo {author}
  {\bibfnamefont {L.}~\bibnamefont {Kuzmin}},\ and\ \bibinfo {author}
  {\bibfnamefont {E.}~\bibnamefont {Il’Ichev}},\ }\bibfield  {title}
  {\bibinfo {title} {Towards a microwave single-photon counter for searching
  axions},\ }\href {https://doi.org/10.1038/s41534-022-00569-5} {\bibfield
  {journal} {\bibinfo  {journal} {npj Quantum Inf.}\ }\textbf {\bibinfo
  {volume} {8}},\ \bibinfo {pages} {61} (\bibinfo {year}
  {2022}{\natexlab{a}})}\BibitemShut {NoStop}%
\bibitem [{\citenamefont {Oelsner}\ \emph {et~al.}(2013)\citenamefont
  {Oelsner}, \citenamefont {Revin}, \citenamefont {Il'ichev}, \citenamefont
  {Pankratov}, \citenamefont {Meyer}, \citenamefont {Grönberg}, \citenamefont
  {Hassel},\ and\ \citenamefont {Kuzmin}}]{JJph4}%
  \BibitemOpen
  \bibfield  {author} {\bibinfo {author} {\bibfnamefont {G.}~\bibnamefont
  {Oelsner}}, \bibinfo {author} {\bibfnamefont {L.~S.}\ \bibnamefont {Revin}},
  \bibinfo {author} {\bibfnamefont {E.}~\bibnamefont {Il'ichev}}, \bibinfo
  {author} {\bibfnamefont {A.~L.}\ \bibnamefont {Pankratov}}, \bibinfo {author}
  {\bibfnamefont {H.-G.}\ \bibnamefont {Meyer}}, \bibinfo {author}
  {\bibfnamefont {L.}~\bibnamefont {Grönberg}}, \bibinfo {author}
  {\bibfnamefont {J.}~\bibnamefont {Hassel}},\ and\ \bibinfo {author}
  {\bibfnamefont {L.~S.}\ \bibnamefont {Kuzmin}},\ }\bibfield  {title}
  {\bibinfo {title} {Underdamped josephson junction as a switching current
  detector},\ }\href {https://doi.org/10.1063/1.4824308} {\bibfield  {journal}
  {\bibinfo  {journal} {Appl. Phys. Lett.}\ }\textbf {\bibinfo {volume}
  {103}},\ \bibinfo {pages} {142605} (\bibinfo {year} {2013})}\BibitemShut
  {NoStop}%
\bibitem [{\citenamefont {Pavlovskiy}\ \emph {et~al.}(2020)\citenamefont
  {Pavlovskiy}, \citenamefont {Gundareva}, \citenamefont {Volkov},\ and\
  \citenamefont {Divin}}]{JJph5}%
  \BibitemOpen
  \bibfield  {author} {\bibinfo {author} {\bibfnamefont {V.~V.}\ \bibnamefont
  {Pavlovskiy}}, \bibinfo {author} {\bibfnamefont {I.~I.}\ \bibnamefont
  {Gundareva}}, \bibinfo {author} {\bibfnamefont {O.~Y.}\ \bibnamefont
  {Volkov}},\ and\ \bibinfo {author} {\bibfnamefont {Y.~Y.}\ \bibnamefont
  {Divin}},\ }\bibfield  {title} {\bibinfo {title} {Wideband detection of
  electromagnetic signals by high-$t_c$ josephson junctions with comparable
  josephson and thermal energies},\ }\href {https://doi.org/10.1063/1.5142400}
  {\bibfield  {journal} {\bibinfo  {journal} {Appl. Phys. Lett.}\ }\textbf
  {\bibinfo {volume} {116}},\ \bibinfo {pages} {082601} (\bibinfo {year}
  {2020})}\BibitemShut {NoStop}%
\bibitem [{\citenamefont {Walsh}\ \emph {et~al.}(2017)\citenamefont {Walsh},
  \citenamefont {Efetov}, \citenamefont {Lee}, \citenamefont {Heuck},
  \citenamefont {Crossno}, \citenamefont {Ohki}, \citenamefont {Kim},
  \citenamefont {Englund},\ and\ \citenamefont
  {Fong}}]{PhysRevApplied.8.024022}%
  \BibitemOpen
  \bibfield  {author} {\bibinfo {author} {\bibfnamefont {E.~D.}\ \bibnamefont
  {Walsh}}, \bibinfo {author} {\bibfnamefont {D.~K.}\ \bibnamefont {Efetov}},
  \bibinfo {author} {\bibfnamefont {G.-H.}\ \bibnamefont {Lee}}, \bibinfo
  {author} {\bibfnamefont {M.}~\bibnamefont {Heuck}}, \bibinfo {author}
  {\bibfnamefont {J.}~\bibnamefont {Crossno}}, \bibinfo {author} {\bibfnamefont
  {T.~A.}\ \bibnamefont {Ohki}}, \bibinfo {author} {\bibfnamefont
  {P.}~\bibnamefont {Kim}}, \bibinfo {author} {\bibfnamefont {D.}~\bibnamefont
  {Englund}},\ and\ \bibinfo {author} {\bibfnamefont {K.~C.}\ \bibnamefont
  {Fong}},\ }\bibfield  {title} {\bibinfo {title} {Graphene-based
  josephson-junction single-photon detector},\ }\href
  {https://doi.org/10.1103/PhysRevApplied.8.024022} {\bibfield  {journal}
  {\bibinfo  {journal} {Phys. Rev. Appl.}\ }\textbf {\bibinfo {volume} {8}},\
  \bibinfo {pages} {024022} (\bibinfo {year} {2017})}\BibitemShut {NoStop}%
\bibitem [{\citenamefont {Lee}\ \emph {et~al.}(2020)\citenamefont {Lee},
  \citenamefont {Efetov}, \citenamefont {Jung}, \citenamefont {Ranzani},
  \citenamefont {Walsh}, \citenamefont {Ohki}, \citenamefont {Taniguchi},
  \citenamefont {Watanabe}, \citenamefont {Kim}, \citenamefont {Englund},\ and\
  \citenamefont {Fong}}]{graphene_2020}%
  \BibitemOpen
  \bibfield  {author} {\bibinfo {author} {\bibfnamefont {G.-H.}\ \bibnamefont
  {Lee}}, \bibinfo {author} {\bibfnamefont {D.~K.}\ \bibnamefont {Efetov}},
  \bibinfo {author} {\bibfnamefont {W.}~\bibnamefont {Jung}}, \bibinfo {author}
  {\bibfnamefont {L.}~\bibnamefont {Ranzani}}, \bibinfo {author} {\bibfnamefont
  {E.~D.}\ \bibnamefont {Walsh}}, \bibinfo {author} {\bibfnamefont {T.~A.}\
  \bibnamefont {Ohki}}, \bibinfo {author} {\bibfnamefont {T.}~\bibnamefont
  {Taniguchi}}, \bibinfo {author} {\bibfnamefont {K.}~\bibnamefont {Watanabe}},
  \bibinfo {author} {\bibfnamefont {P.}~\bibnamefont {Kim}}, \bibinfo {author}
  {\bibfnamefont {D.}~\bibnamefont {Englund}},\ and\ \bibinfo {author}
  {\bibfnamefont {K.~C.}\ \bibnamefont {Fong}},\ }\bibfield  {title} {\bibinfo
  {title} {Graphene-based josephson junction microwave bolometer},\ }\href
  {https://doi.org/10.1038/s41586-020-2752-4} {\bibfield  {journal} {\bibinfo
  {journal} {Nature}\ }\textbf {\bibinfo {volume} {586}},\ \bibinfo {pages}
  {42} (\bibinfo {year} {2020})}\BibitemShut {NoStop}%
\bibitem [{\citenamefont {Chen}\ \emph {et~al.}(1988)\citenamefont {Chen},
  \citenamefont {Fisher},\ and\ \citenamefont {Leggett}}]{retrapping}%
  \BibitemOpen
  \bibfield  {author} {\bibinfo {author} {\bibfnamefont {Y.~C.}\ \bibnamefont
  {Chen}}, \bibinfo {author} {\bibfnamefont {M.~P.~A.}\ \bibnamefont
  {Fisher}},\ and\ \bibinfo {author} {\bibfnamefont {A.~J.}\ \bibnamefont
  {Leggett}},\ }\bibfield  {title} {\bibinfo {title} {The return of a
  hysteretic josephson junction to the zero‐voltage state: I‐v
  characteristic and quantum retrapping},\ }\href
  {https://doi.org/10.1063/1.341527} {\bibfield  {journal} {\bibinfo  {journal}
  {J. Appl. Phys.}\ }\textbf {\bibinfo {volume} {64}},\ \bibinfo {pages} {3119}
  (\bibinfo {year} {1988})}\BibitemShut {NoStop}%
\bibitem [{\citenamefont {Pankratov}\ \emph
  {et~al.}(2022{\natexlab{b}})\citenamefont {Pankratov}, \citenamefont
  {Gordeeva}, \citenamefont {Revin}, \citenamefont {Ladeynov}, \citenamefont
  {Yablokov},\ and\ \citenamefont {Kuzmin}}]{Pankratov2022}%
  \BibitemOpen
  \bibfield  {author} {\bibinfo {author} {\bibfnamefont {A.~L.}\ \bibnamefont
  {Pankratov}}, \bibinfo {author} {\bibfnamefont {A.~V.}\ \bibnamefont
  {Gordeeva}}, \bibinfo {author} {\bibfnamefont {L.~S.}\ \bibnamefont {Revin}},
  \bibinfo {author} {\bibfnamefont {D.~A.}\ \bibnamefont {Ladeynov}}, \bibinfo
  {author} {\bibfnamefont {A.~A.}\ \bibnamefont {Yablokov}},\ and\ \bibinfo
  {author} {\bibfnamefont {L.~S.}\ \bibnamefont {Kuzmin}},\ }\bibfield  {title}
  {\bibinfo {title} {Approaching microwave photon sensitivity with al josephson
  junctions},\ }\href {https://doi.org/10.3762/bjnano.13.50} {\bibfield
  {journal} {\bibinfo  {journal} {Beilstein J. Nanotechnol.}\ }\textbf
  {\bibinfo {volume} {13}},\ \bibinfo {pages} {582} (\bibinfo {year}
  {2022}{\natexlab{b}})}\BibitemShut {NoStop}%
\bibitem [{\citenamefont {Guarcello}\ \emph {et~al.}(2019)\citenamefont
  {Guarcello}, \citenamefont {Valenti}, \citenamefont {Spagnolo}, \citenamefont
  {Pierro},\ and\ \citenamefont {Filatrella}}]{JJph7}%
  \BibitemOpen
  \bibfield  {author} {\bibinfo {author} {\bibfnamefont {C.}~\bibnamefont
  {Guarcello}}, \bibinfo {author} {\bibfnamefont {D.}~\bibnamefont {Valenti}},
  \bibinfo {author} {\bibfnamefont {B.}~\bibnamefont {Spagnolo}}, \bibinfo
  {author} {\bibfnamefont {V.}~\bibnamefont {Pierro}},\ and\ \bibinfo {author}
  {\bibfnamefont {G.}~\bibnamefont {Filatrella}},\ }\bibfield  {title}
  {\bibinfo {title} {Josephson-based threshold detector for l\'evy-distributed
  current fluctuations},\ }\href
  {https://doi.org/10.1103/PhysRevApplied.11.044078} {\bibfield  {journal}
  {\bibinfo  {journal} {Phys. Rev. Appl.}\ }\textbf {\bibinfo {volume} {11}},\
  \bibinfo {pages} {044078} (\bibinfo {year} {2019})}\BibitemShut {NoStop}%
\bibitem [{\citenamefont {Fulton}\ and\ \citenamefont
  {Dunkleberger}(1974)}]{PhysRevB.9.4760}%
  \BibitemOpen
  \bibfield  {author} {\bibinfo {author} {\bibfnamefont {T.~A.}\ \bibnamefont
  {Fulton}}\ and\ \bibinfo {author} {\bibfnamefont {L.~N.}\ \bibnamefont
  {Dunkleberger}},\ }\bibfield  {title} {\bibinfo {title} {Lifetime of the
  zero-voltage state in josephson tunnel junctions},\ }\href
  {https://doi.org/10.1103/PhysRevB.9.4760} {\bibfield  {journal} {\bibinfo
  {journal} {Phys. Rev. B}\ }\textbf {\bibinfo {volume} {9}},\ \bibinfo {pages}
  {4760} (\bibinfo {year} {1974})}\BibitemShut {NoStop}%
\bibitem [{\citenamefont {Kurkij\"arvi}(1972)}]{noise1}%
  \BibitemOpen
  \bibfield  {author} {\bibinfo {author} {\bibfnamefont {J.}~\bibnamefont
  {Kurkij\"arvi}},\ }\bibfield  {title} {\bibinfo {title} {Intrinsic
  fluctuations in a superconducting ring closed with a josephson junction},\
  }\href {https://doi.org/10.1103/PhysRevB.6.832} {\bibfield  {journal}
  {\bibinfo  {journal} {Phys. Rev. B}\ }\textbf {\bibinfo {volume} {6}},\
  \bibinfo {pages} {832} (\bibinfo {year} {1972})}\BibitemShut {NoStop}%
\bibitem [{\citenamefont {Tobiska}\ and\ \citenamefont
  {Nazarov}(2004)}]{PhysRevLett.93.106801}%
  \BibitemOpen
  \bibfield  {author} {\bibinfo {author} {\bibfnamefont {J.}~\bibnamefont
  {Tobiska}}\ and\ \bibinfo {author} {\bibfnamefont {Y.~V.}\ \bibnamefont
  {Nazarov}},\ }\bibfield  {title} {\bibinfo {title} {Josephson junctions as
  threshold detectors for full counting statistics},\ }\href
  {https://doi.org/10.1103/PhysRevLett.93.106801} {\bibfield  {journal}
  {\bibinfo  {journal} {Phys. Rev. Lett.}\ }\textbf {\bibinfo {volume} {93}},\
  \bibinfo {pages} {106801} (\bibinfo {year} {2004})}\BibitemShut {NoStop}%
\bibitem [{\citenamefont {He}\ \emph {et~al.}(2025)\citenamefont {He},
  \citenamefont {Ouyang}, \citenamefont {Chai}, \citenamefont {Chang},
  \citenamefont {He},\ and\ \citenamefont {Wei}}]{our2}%
  \BibitemOpen
  \bibfield  {author} {\bibinfo {author} {\bibfnamefont {J.-X.}\ \bibnamefont
  {He}}, \bibinfo {author} {\bibfnamefont {P.-H.}\ \bibnamefont {Ouyang}},
  \bibinfo {author} {\bibfnamefont {Y.-Q.}\ \bibnamefont {Chai}}, \bibinfo
  {author} {\bibfnamefont {H.}~\bibnamefont {Chang}}, \bibinfo {author}
  {\bibfnamefont {Q.}~\bibnamefont {He}},\ and\ \bibinfo {author}
  {\bibfnamefont {L.-F.}\ \bibnamefont {Wei}},\ }\bibfield  {title} {\bibinfo
  {title} {Experimental demonstrations of josephson threshold detectors for
  broadband microwave photons detection},\ }\href
  {https://doi.org/10.1063/5.0259463} {\bibfield  {journal} {\bibinfo
  {journal} {Appl. Phys. Lett.}\ }\textbf {\bibinfo {volume} {126}},\ \bibinfo
  {pages} {202601} (\bibinfo {year} {2025})}\BibitemShut {NoStop}%
\bibitem [{\citenamefont {Cheng}\ \emph {et~al.}(2018)\citenamefont {Cheng},
  \citenamefont {Cirillo}, \citenamefont {Salina},\ and\ \citenamefont
  {Gr\o{}nbech-Jensen}}]{SCD0}%
  \BibitemOpen
  \bibfield  {author} {\bibinfo {author} {\bibfnamefont {C.}~\bibnamefont
  {Cheng}}, \bibinfo {author} {\bibfnamefont {M.}~\bibnamefont {Cirillo}},
  \bibinfo {author} {\bibfnamefont {G.}~\bibnamefont {Salina}},\ and\ \bibinfo
  {author} {\bibfnamefont {N.}~\bibnamefont {Gr\o{}nbech-Jensen}},\ }\bibfield
  {title} {\bibinfo {title} {Nonequilibrium transient phenomena in the
  washboard potential},\ }\href {https://doi.org/10.1103/PhysRevE.98.012140}
  {\bibfield  {journal} {\bibinfo  {journal} {Phys. Rev. E}\ }\textbf {\bibinfo
  {volume} {98}},\ \bibinfo {pages} {012140} (\bibinfo {year}
  {2018})}\BibitemShut {NoStop}%
\bibitem [{\citenamefont {Barone}\ \emph {et~al.}(1985)\citenamefont {Barone},
  \citenamefont {Cristiano},\ and\ \citenamefont
  {Silvestrini}}]{nonequilibrium1}%
  \BibitemOpen
  \bibfield  {author} {\bibinfo {author} {\bibfnamefont {A.}~\bibnamefont
  {Barone}}, \bibinfo {author} {\bibfnamefont {R.}~\bibnamefont {Cristiano}},\
  and\ \bibinfo {author} {\bibfnamefont {P.}~\bibnamefont {Silvestrini}},\
  }\bibfield  {title} {\bibinfo {title} {Supercurrent decay in underdamped
  josephson junctions: Nonstationary case},\ }\href
  {https://doi.org/10.1063/1.335597} {\bibfield  {journal} {\bibinfo  {journal}
  {J. Appl. Phys.}\ }\textbf {\bibinfo {volume} {58}},\ \bibinfo {pages} {3822}
  (\bibinfo {year} {1985})}\BibitemShut {NoStop}%
\bibitem [{\citenamefont {Silvestrini}\ \emph {et~al.}(1997)\citenamefont
  {Silvestrini}, \citenamefont {Palmieri}, \citenamefont {Ruggiero},\ and\
  \citenamefont {Russo}}]{nonequilibrium2}%
  \BibitemOpen
  \bibfield  {author} {\bibinfo {author} {\bibfnamefont {P.}~\bibnamefont
  {Silvestrini}}, \bibinfo {author} {\bibfnamefont {V.~G.}\ \bibnamefont
  {Palmieri}}, \bibinfo {author} {\bibfnamefont {B.}~\bibnamefont {Ruggiero}},\
  and\ \bibinfo {author} {\bibfnamefont {M.}~\bibnamefont {Russo}},\ }\bibfield
   {title} {\bibinfo {title} {Observation of energy levels quantization in
  underdamped josephson junctions above the classical-quantum regime crossover
  temperature},\ }\href {https://doi.org/10.1103/PhysRevLett.79.3046}
  {\bibfield  {journal} {\bibinfo  {journal} {Phys. Rev. Lett.}\ }\textbf
  {\bibinfo {volume} {79}},\ \bibinfo {pages} {3046} (\bibinfo {year}
  {1997})}\BibitemShut {NoStop}%
\bibitem [{\citenamefont {Blackburn}\ \emph {et~al.}(2016)\citenamefont
  {Blackburn}, \citenamefont {Cirillo},\ and\ \citenamefont
  {Grønbech-Jensen}}]{RCSJ3}%
  \BibitemOpen
  \bibfield  {author} {\bibinfo {author} {\bibfnamefont {J.~A.}\ \bibnamefont
  {Blackburn}}, \bibinfo {author} {\bibfnamefont {M.}~\bibnamefont {Cirillo}},\
  and\ \bibinfo {author} {\bibfnamefont {N.}~\bibnamefont {Grønbech-Jensen}},\
  }\bibfield  {title} {\bibinfo {title} {A survey of classical and quantum
  interpretations of experiments on josephson junctions at very low
  temperatures},\ }\href
  {https://doi.org/https://doi.org/10.1016/j.physrep.2015.10.010} {\bibfield
  {journal} {\bibinfo  {journal} {Phys. Rep.}\ }\textbf {\bibinfo {volume}
  {611}},\ \bibinfo {pages} {1} (\bibinfo {year} {2016})}\BibitemShut {NoStop}%
\bibitem [{\citenamefont {Falco}\ \emph {et~al.}(1974)\citenamefont {Falco},
  \citenamefont {Parker}, \citenamefont {Trullinger},\ and\ \citenamefont
  {Hansma}}]{thermal}%
  \BibitemOpen
  \bibfield  {author} {\bibinfo {author} {\bibfnamefont {C.~M.}\ \bibnamefont
  {Falco}}, \bibinfo {author} {\bibfnamefont {W.~H.}\ \bibnamefont {Parker}},
  \bibinfo {author} {\bibfnamefont {S.~E.}\ \bibnamefont {Trullinger}},\ and\
  \bibinfo {author} {\bibfnamefont {P.~K.}\ \bibnamefont {Hansma}},\ }\bibfield
   {title} {\bibinfo {title} {Effect of thermal noise on current-voltage
  characteristics of josephson junctions},\ }\href
  {https://doi.org/10.1103/PhysRevB.10.1865} {\bibfield  {journal} {\bibinfo
  {journal} {Phys. Rev. B}\ }\textbf {\bibinfo {volume} {10}},\ \bibinfo
  {pages} {1865} (\bibinfo {year} {1974})}\BibitemShut {NoStop}%
\bibitem [{\citenamefont {Kramers}(1940)}]{noise2}%
  \BibitemOpen
  \bibfield  {author} {\bibinfo {author} {\bibfnamefont {H.~A.}\ \bibnamefont
  {Kramers}},\ }\bibfield  {title} {\bibinfo {title} {Brownian motion in a
  field of force and the diffusion model of chemical reactions},\ }\href
  {https://doi.org/https://doi.org/10.1016/S0031-8914(40)90098-2} {\bibfield
  {journal} {\bibinfo  {journal} {Physica}\ }\textbf {\bibinfo {volume} {7}},\
  \bibinfo {pages} {284} (\bibinfo {year} {1940})}\BibitemShut {NoStop}%
\bibitem [{\citenamefont {Ruggiero}\ \emph {et~al.}(1998)\citenamefont
  {Ruggiero}, \citenamefont {Granata}, \citenamefont {Palmieri}, \citenamefont
  {Esposito}, \citenamefont {Russo},\ and\ \citenamefont
  {Silvestrini}}]{SCD_eq}%
  \BibitemOpen
  \bibfield  {author} {\bibinfo {author} {\bibfnamefont {B.}~\bibnamefont
  {Ruggiero}}, \bibinfo {author} {\bibfnamefont {C.}~\bibnamefont {Granata}},
  \bibinfo {author} {\bibfnamefont {V.~G.}\ \bibnamefont {Palmieri}}, \bibinfo
  {author} {\bibfnamefont {A.}~\bibnamefont {Esposito}}, \bibinfo {author}
  {\bibfnamefont {M.}~\bibnamefont {Russo}},\ and\ \bibinfo {author}
  {\bibfnamefont {P.}~\bibnamefont {Silvestrini}},\ }\bibfield  {title}
  {\bibinfo {title} {Supercurrent decay in extremely underdamped josephson
  junctions},\ }\href {https://doi.org/10.1103/PhysRevB.57.134} {\bibfield
  {journal} {\bibinfo  {journal} {Phys. Rev. B}\ }\textbf {\bibinfo {volume}
  {57}},\ \bibinfo {pages} {134} (\bibinfo {year} {1998})}\BibitemShut
  {NoStop}%
\bibitem [{\citenamefont {Caldeira}\ and\ \citenamefont
  {Leggett}(1981)}]{Leggett}%
  \BibitemOpen
  \bibfield  {author} {\bibinfo {author} {\bibfnamefont {A.~O.}\ \bibnamefont
  {Caldeira}}\ and\ \bibinfo {author} {\bibfnamefont {A.~J.}\ \bibnamefont
  {Leggett}},\ }\bibfield  {title} {\bibinfo {title} {Influence of dissipation
  on quantum tunneling in macroscopic systems},\ }\href
  {https://doi.org/10.1103/PhysRevLett.46.211} {\bibfield  {journal} {\bibinfo
  {journal} {Phys. Rev. Lett.}\ }\textbf {\bibinfo {volume} {46}},\ \bibinfo
  {pages} {211} (\bibinfo {year} {1981})}\BibitemShut {NoStop}%
\bibitem [{\citenamefont {Muschelli~III}(2020)}]{ROC2}%
  \BibitemOpen
  \bibfield  {author} {\bibinfo {author} {\bibfnamefont {J.}~\bibnamefont
  {Muschelli~III}},\ }\bibfield  {title} {\bibinfo {title} {Roc and auc with a
  binary predictor: a potentially misleading metric},\ }\href
  {https://doi.org/https://doi.org/10.1007/s00357-019-09345-1} {\bibfield
  {journal} {\bibinfo  {journal} {J. Classif.}\ }\textbf {\bibinfo {volume}
  {37}},\ \bibinfo {pages} {696} (\bibinfo {year} {2020})}\BibitemShut
  {NoStop}%
\bibitem [{\citenamefont {Yang}\ \emph {et~al.}(2021)\citenamefont {Yang},
  \citenamefont {Luo}, \citenamefont {Fan}, \citenamefont {Sha}, \citenamefont
  {Xiao},\ and\ \citenamefont {Cui}}]{ROC3}%
  \BibitemOpen
  \bibfield  {author} {\bibinfo {author} {\bibfnamefont {C.}~\bibnamefont
  {Yang}}, \bibinfo {author} {\bibfnamefont {X.}~\bibnamefont {Luo}}, \bibinfo
  {author} {\bibfnamefont {L.}~\bibnamefont {Fan}}, \bibinfo {author}
  {\bibfnamefont {W.}~\bibnamefont {Sha}}, \bibinfo {author} {\bibfnamefont
  {H.}~\bibnamefont {Xiao}},\ and\ \bibinfo {author} {\bibfnamefont
  {H.}~\bibnamefont {Cui}},\ }\bibfield  {title} {\bibinfo {title} {Performance
  of interferon-gamma release assays in the diagnosis of nontuberculous
  mycobacterial diseases—a retrospective survey from 2011 to 2019},\ }\href
  {https://doi.org/10.3389/fcimb.2020.571230} {\bibfield  {journal} {\bibinfo
  {journal} {Front. Cell. Infect. Microbiol.}\ }\textbf {\bibinfo {volume}
  {10}},\ \bibinfo {pages} {571230} (\bibinfo {year} {2021})}\BibitemShut
  {NoStop}%
\bibitem [{\citenamefont {Bishop}\ and\ \citenamefont
  {Nasrabadi}(2006)}]{classify}%
  \BibitemOpen
  \bibfield  {author} {\bibinfo {author} {\bibfnamefont {C.~M.}\ \bibnamefont
  {Bishop}}\ and\ \bibinfo {author} {\bibfnamefont {N.~M.}\ \bibnamefont
  {Nasrabadi}},\ }\href@noop {} {\emph {\bibinfo {title} {Pattern recognition
  and machine learning}}},\ Vol.~\bibinfo {volume} {4}\ (\bibinfo  {publisher}
  {Springer},\ \bibinfo {year} {2006})\BibitemShut {NoStop}%
\bibitem [{\citenamefont {Krasnov}(2024)}]{PhysRevApplied.22.024015}%
  \BibitemOpen
  \bibfield  {author} {\bibinfo {author} {\bibfnamefont {V.~M.}\ \bibnamefont
  {Krasnov}},\ }\bibfield  {title} {\bibinfo {title} {Resonant switching
  current detector based on underdamped josephson junctions},\ }\href
  {https://doi.org/10.1103/PhysRevApplied.22.024015} {\bibfield  {journal}
  {\bibinfo  {journal} {Phys. Rev. Appl.}\ }\textbf {\bibinfo {volume} {22}},\
  \bibinfo {pages} {024015} (\bibinfo {year} {2024})}\BibitemShut {NoStop}%
\bibitem [{\citenamefont {Revin}\ \emph {et~al.}(2023)\citenamefont {Revin},
  \citenamefont {Ladeynov}, \citenamefont {Gordeeva},\ and\ \citenamefont
  {Pankratov}}]{pulse}%
  \BibitemOpen
  \bibfield  {author} {\bibinfo {author} {\bibfnamefont {L.}~\bibnamefont
  {Revin}}, \bibinfo {author} {\bibfnamefont {D.}~\bibnamefont {Ladeynov}},
  \bibinfo {author} {\bibfnamefont {A.}~\bibnamefont {Gordeeva}},\ and\
  \bibinfo {author} {\bibfnamefont {A.}~\bibnamefont {Pankratov}},\ }\bibfield
  {title} {\bibinfo {title} {Response of a josephson junction to a current
  pulse with the energy of a microwave photon},\ }\href
  {https://doi.org/10.61011/PSS.2023.07.56389.30H} {\bibfield  {journal}
  {\bibinfo  {journal} {Phys. Solid State}\ }\textbf {\bibinfo {volume} {65}},\
  \bibinfo {pages} {1043} (\bibinfo {year} {2023})}\BibitemShut {NoStop}%
\bibitem [{\citenamefont {Ouyang}\ \emph {et~al.}(2024)\citenamefont {Ouyang},
  \citenamefont {He}, \citenamefont {Wang}, \citenamefont {Chai}, \citenamefont
  {He}, \citenamefont {Chang},\ and\ \citenamefont {Wei}}]{our3}%
  \BibitemOpen
  \bibfield  {author} {\bibinfo {author} {\bibfnamefont {P.~H.}\ \bibnamefont
  {Ouyang}}, \bibinfo {author} {\bibfnamefont {S.~R.}\ \bibnamefont {He}},
  \bibinfo {author} {\bibfnamefont {Y.~Z.}\ \bibnamefont {Wang}}, \bibinfo
  {author} {\bibfnamefont {Y.~Q.}\ \bibnamefont {Chai}}, \bibinfo {author}
  {\bibfnamefont {J.~X.}\ \bibnamefont {He}}, \bibinfo {author} {\bibfnamefont
  {H.}~\bibnamefont {Chang}},\ and\ \bibinfo {author} {\bibfnamefont {L.~F.}\
  \bibnamefont {Wei}},\ }\bibfield  {title} {\bibinfo {title} {Experimental
  evidence for a current-biased josephson junction acting as either a
  macroscopic boson or fermion},\ }\href
  {https://doi.org/10.1103/PhysRevResearch.6.013236} {\bibfield  {journal}
  {\bibinfo  {journal} {Phys. Rev. Res.}\ }\textbf {\bibinfo {volume} {6}},\
  \bibinfo {pages} {013236} (\bibinfo {year} {2024})}\BibitemShut {NoStop}%
\bibitem [{\citenamefont {Ravanne}\ \emph {et~al.}(2022)\citenamefont
  {Ravanne}, \citenamefont {Then}, \citenamefont {Su},\ and\ \citenamefont
  {Hijazin}}]{9769720}%
  \BibitemOpen
  \bibfield  {author} {\bibinfo {author} {\bibfnamefont {J.~G.}\ \bibnamefont
  {Ravanne}}, \bibinfo {author} {\bibfnamefont {Y.~L.}\ \bibnamefont {Then}},
  \bibinfo {author} {\bibfnamefont {H.~T.}\ \bibnamefont {Su}},\ and\ \bibinfo
  {author} {\bibfnamefont {I.}~\bibnamefont {Hijazin}},\ }\bibfield  {title}
  {\bibinfo {title} {Microwave power detectors in different cmos design
  architectures: A review},\ }\href {https://doi.org/10.1109/MMM.2022.3155033}
  {\bibfield  {journal} {\bibinfo  {journal} {IEEE Microwave}\ }\textbf
  {\bibinfo {volume} {23}},\ \bibinfo {pages} {76} (\bibinfo {year}
  {2022})}\BibitemShut {NoStop}%
\end{thebibliography}%
\end{document}